\documentclass[12pt]{iopart}
\usepackage{mathrsfs} 
\expandafter\let\csname equation*\endcsname\relax
\expandafter\let\csname endequation*\endcsname\relax
\usepackage{amsmath}
\usepackage[version=4]{mhchem}
\usepackage{amsfonts} 
\usepackage{amsmath,amssymb}
\usepackage{graphicx}
\usepackage{subfig}
\usepackage{float}
\usepackage[table]{xcolor}
\usepackage{colortbl}
\usepackage{tikz}
\usepackage{pgfplots}
\usepackage{bm}
\usepackage{multirow}
\usepackage[explicit]{titlesec}
\usepackage[most]{tcolorbox}
\usepackage{braket}
\usepackage{soul}

\usepackage{iopams}  
\usepackage{indentfirst}
\usepackage{hyperref}
\hypersetup{
    colorlinks=true,
    linkcolor=blue,
    filecolor=blue,      
    urlcolor=blue,
    citecolor=blue}
\pgfplotsset{compat=1.18}

\begin{document}

\title{The Kondo effect in the quantum $XX$ spin chain}

\author{Pradip Kattel, Yicheng Tang, J.~H.~Pixley and  Natan Andrei}

\address{ Department of Physics and Astronomy, Center for Materials Theory, Rutgers University, Piscataway, New Jersey 08854, USA}
\ead{pradip.kattel@rutgers.edu}
\vspace{10pt}
\begin{indented}
\item[]January 2024
\end{indented}

\begin{abstract}
 We investigate the boundary phenomena that arise in a finite-size $XX$ spin chain interacting through an $XX$ interaction with a spin$-\frac{1}{2}$ impurity located at its edge. Upon Jordan-Wigner transformation, the model is described by a quadratic Fermionic Hamiltonian. Our work displays, within this ostensibly simple model, the emergence of the Kondo effect, a quintessential hallmark of strongly correlated physics. We also show how the Kondo cloud shrinks and turns into a single particle bound state as the impurity coupling increases beyond a critical value. In more detail, using both \textit{Bethe Ansatz} and \textit{exact diagonalization} (ED) techniques, we show that the local moment of the impurity is screened by different mechanisms depending on the ratio of the boundary and bulk coupling $\frac{J_{\mathrm{imp}}}{J}$. When the ratio falls below the critical value $\sqrt{2}$, the impurity is screened via the multiparticle Kondo effect. However, when
 the ratio between the coupling exceeds the critical value   $\sqrt{2}$ an exponentially localized bound mode is formed at the impurity site which screens the spin of the impurity in the ground state. We show that the boundary phase transition is reflected in local ground state properties by calculating the spinon density of states, the magnetization at the impurity site in the presence of a global magnetic field, and the finite temperature susceptibility of the impurity. We find that the spinon density of states in the Kondo phase has the characteristic Lorentzian peak that moves from the Fermi level to the maximum energy of the spinon as the impurity coupling is increased and becomes a localized bound mode in the bound mode phase. Moreover, the impurity magnetization and the finite temperature impurity susceptibility behave differently in the two phases. When the boundary coupling $J_{\mathrm{imp}}$ exceeds the critical value $\sqrt{2}J$, the model is no longer boundary conformal invariant as a massive bound mode appears at the impurity site. 
\end{abstract}


\section{Introduction}
The antiferromagnetic Kondo effect is a quintessential example of a strongly correlated phenomenon. It was first observed in the resistivity of a metal with a dilute concentration of magnetic impurities. This results,  among many other effects, in the spin-$\frac12$ impurity being screened at low temperature while behaving as a free spin at high temperatures \cite{hewson1997kondo, kondo2012physics}.
Theoretically, this phenomenon was interpreted as a smooth cross-over from weak coupling at high temperatures to strong coupling at low temperatures mediated by non-perturbative spin-flip processes\cite{anderson2018poor, wilson1975renormalization}.

In the conventional Kondo problem, considering one impurity at energy scales small compared to the Fermi energy one can describe the system by an effective one-dimensional free fermion gas with a linearized spectrum perturbed by a marginally relevant impurity (defect) operator \cite{wilson1975renormalization, hewson1997kondo,kondo2012physics}. This description makes the problem amenable to various non-perturbative treatments, the Wilson numerical RG, the \textit{Bethe Ansatz}, and the conformal field theory, all of which overcome the failure of perturbative calculations
\cite{wilson1975renormalization,nozieres1974fermi,andrei1980diagonalization,wiegmann1981exact,affleck1991kondo}. Kondo behavior also arises in other contexts, in particular in spin chains, when an antiferromagnetic spin chain interacts with a spin impurity in the bulk \cite{andrei1984heisenberg} or at its boundary\cite{laflorencie2008kondo, wang1997exact, zvyagin1997magnetic, kattel2023kondo}. In this paper, we demonstrate the occurrence of a similar effect in the $XX$ spin chain with a boundary impurity, which can also be described by a non-interacting fermionic Hamiltonian. This model shows Kondo behavior in physical quantities such as the signature Lorentzian-like peak in the impurity density of states, a smooth crossover of local impurity magnetization, and a characteristic susceptibility that is finite at low temperature but falls off as $\frac{1}{T}$ at high temperature, when the boundary coupling is sufficiently weaker than the bulk coupling. However, when the boundary coupling is larger than $\sqrt{2}$ bulk coupling 
, the impurity opens a boundary gap due to an appearance of localized massive mode at the boundary. It is remarkable to find the Kondo effect and a distinct bound mode phase in {this quantum spin model that is mapable to
a quadratic, free fermion model}.

Focusing on the effective one-dimensional theory with a linearized energy spectrum near the Fermi energy, due to spin-charge separation
the charge part of the  fermion decouples and only the spin fluctuations 
couple to the impurity operator, thereby showing that the Kondo effect can be described by the spin sector of the electrons that interact with a spin defect. This manifests itself in the spectrum that consists of decoupled charge {\it{holons}}  and spin-$\frac12$ {\it{spinons} }coupled to the impurity\cite{andrei1980diagonalization}, or by nonabelian bosonization that allows the separation of charge and spin degree of freedom in the Hamiltonian \cite{affleck1995conformal, gogolin2004bosonization}. In the low energy sector, similar considerations apply to the system we are considering in this paper, the $XX$ spin chain with a point defect at the boundary, described by the Hamiltonian
\begin{equation}
   \mathcal{H} = \sum_{j=1}^{N-1} J (\sigma_j^x \sigma_{j+1}^x + \sigma_j^y \sigma_{j+1}^y) + J_{\mathrm{imp}} (\sigma_1^x \sigma_{0}^x + \sigma_1^y \sigma_{0}^y).
   \label{hammodel}
\end{equation}
Here $\vec{\sigma}_j$ are the Pauli matrices that act on the local two-dimensional Hilbert space of each quantum spin variable and $\vec{\sigma}_0$ is the impurity spin operator located at the left edge of the spin chain and interacting with the first spin in the spin chain. Upon fermionizing the spin variable via the Jordan-Wigner transformation, the bulk describes a gas of free fermions on the lattice (i.e.~the tight-binding model) with a bond defect
\begin{equation}
    \mathcal{H}=\sum_{j=1}^{N-1}J(\psi^\dagger_j \psi_{j+1}+\psi_{j+1}^\dagger \psi_j)+ J_{\mathrm{imp}}(\psi^\dagger_0\psi_1+\psi_1^\dagger \psi_0 ).
\end{equation}
To take the continuum limit, recall the relation between the spin operators and the currents $\vec \sigma_j \sim\frac{1}{2\pi}[\vec{\mathcal{J}}_L)(aj)+\vec{\mathcal{J}}_R)(aj)]+(-1)^j \mathrm{constant}~\vec n(aj)$ where $a$ is the lattice spacing and both uniform currents $\vec{\mathcal{J}}_{R/L}$ and the staggered part $\vec n$ vary slowly at long distance \cite{laflorencie2008kondo}.
With the boundary condition $\vec{\mathcal{J}}_L(x=0)=\vec {\mathcal{J}}_R(x=0)$  such that $\vec {\mathcal J}_R(x)$ can be regarded as the analytic
continuation of $\vec {\mathcal J}_L(x)$ to the negative x-axis and the staggered part $\vec n(0)=\vec{\mathcal{J}}_L(0)=\vec{\mathcal{J}}_R(0)$ at the boundary, the low energy Hamiltonian density  can be written as a perturbed quadratic theory in terms of spin currents \cite{laflorencie2008kondo,giamarchi2003quantum}
\begin{equation}
   \mathcal H\sim \left[\vec{\mathcal J}(x)\right]^2+ g \delta(x)\left(\mathcal J^x S^x+\mathcal J^y S^y\right),
    \label{bosxx}
\end{equation}
where $g\propto\frac{J_{\mathrm{imp}}}{J}$ and we dropped the chirality index. Apart from the lack of the $\mathcal{J}^z S^z$ interaction, the  Hamiltonian in Eq.\eqref{bosxx} is the same as the spin part of the conventional Kondo Hamiltonian \cite{affleck1995conformal} describing the bath of free fermions interacting with a spin impurity.


\begin{figure}[H]
    \centering
    \includegraphics[scale=0.75]{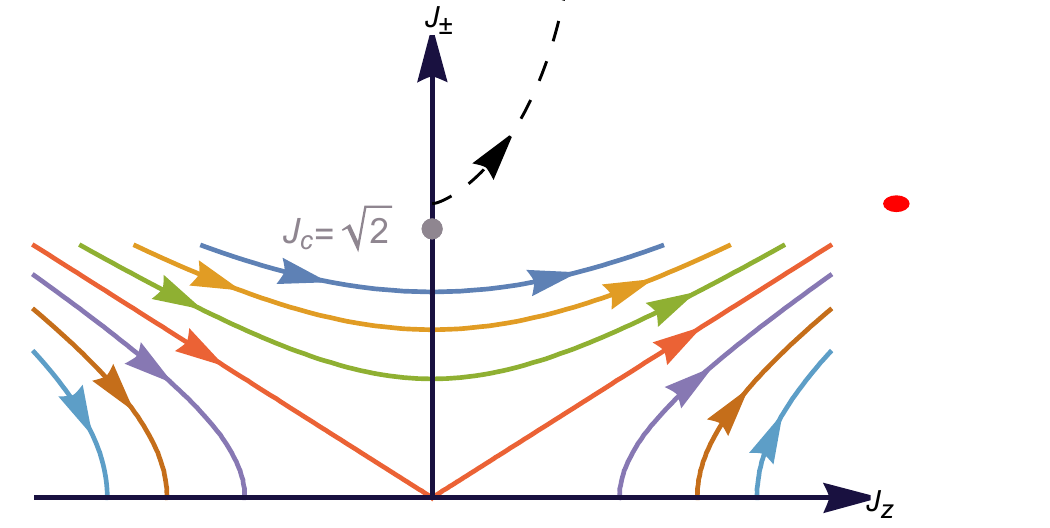}
    \caption{Perturbative renormalization group flow diagram of the anisotropic Kondo impurity model which shows that only when the impurity coupling is ferromagnetic, the flow is towards the trivial fix point but for rest of the parametric regimes, the model flows to the strong coupling fixed points denoted by red circle in the schematic. The low energy field theoretic description of the $XX$-Kondo model considered in this paper also flows to the same coupling fixed point when the impurity coupling is below the critical value of $\sqrt{2}$ times the bulk coupling. But when the boundary coupling exceeds the critical value, the boundary opens a gap and the model is no longer critical.}
    \label{fig:RG-akm}
\end{figure}
{To see how the strong coupling fixed point is retained in Eq.~\eqref{bosxx}, we recall}
 the RG equations for the anisotropic Kondo model with parallel coupling $g^z$ and perpendicular couplings $g^\pm$\cite{kogan2017spin,kogan2018poor}
\begin{equation}
\begin{aligned}
& \frac{\mathrm{d} g_z}{\mathrm{~d} \ln \Lambda}=-2 g_{ \pm}^2+\mathcal{O}\left(g^3\right), \quad  \frac{\mathrm{d} g_{ \pm}}{\mathrm{d} \ln \Lambda}=-2 g_z g_{ \pm}+\mathcal{O}\left(g^3\right).\label{rgfloweqn}
\end{aligned}
\end{equation}
 As shown in Fig.\ref{fig:RG-akm}, both the conventional Kondo model and the $XX-$Kondo model (for small boundary coupling)  are described by  Eq.\eqref{bosxx} and flow to the same strong coupling fix point characterized by the quenching of the local magnetic moment of the impurity. The RG-flow is also instructive to see that the $XX-$Kondo model does not have a Berzinski-Kosterlitz-Thouless transition; it always flows to strong coupling and the irrelevant ferromagnetic Kondo coupling regime is not a part of the model or its effective theory.
 It is also important to note that even though the $z-$ part of the Kondo coupling is absent in the $XX-$Kondo model in the starting theory, as seen from Eq.\eqref{rgfloweqn}, the model flows to the same fixed point in the strong coupling limit as such an interaction is generated in the RG process. Thus, both models have similar physics in the low energy limit but with 
 the several differences e.g. due to the absence of a Kondo coupling in the $z$-direction and the effect of the UV cut-off provided by the lattice spacing.  Here, using Bethe Ansatz, we solve the Hamiltonian, Eq.\eqref{hammodel}, and study it  on all energy scales. We show that for $J_{\mathrm{imp}}<\sqrt{2}$  the model is in a Kondo phase where the impurity is screened by a multi-particle Kondo cloud,  and it undergoes a boundary phase transition between the Kondo phase and a bound-mode phase where the impurity is screened by a single bound mode formed at the impurity site. Our results show that Eq.\eqref{hammodel} does not remain boundary conformal invariant for all values of $J_{\mathrm{imp}}$ at low energy scale. Although the bulk remains critical, there are boundary excitations that are massive when $J_{\mathrm{imp}}>\sqrt{2}J$ which breaks the conformal invariance. 

Quantum impurities at the edge of various interacting models have been studied before by various authors \cite{rylands2020exact,kattel2023kondo,zvyagin1997magnetic,frahm1997open,pasnoori2020kondo,lee1992kondo,furusaki1994kondo,frojdh1995kondo}. 
The model we studied here, which can be mapped to a non-interacting Fermionic problem, nonetheless, has a rich phase diagram that resembles the Kondo-like phase and a distinct bound mode phase.  Recently, we considered the isotropic Heisenberg chain with boundary impurity \cite{kattel2023kondo}, where the similar physics of boundary impurity phenomena occurs with the presence of $\sigma^z_i\sigma^z_{i+1}$ coupling when antiferromagnetic impurity considered is considered. There the phase transition between the Kondo and bound mode occurs at $J_\mathrm{imp}=\frac{4}{3}J$. The sharp contrast is when $J_\mathrm{imp}<0$ where impurity is unscreened or can be screened by bound mode in the case of $XXX$ but for $XX$, the sign of $J_{\mathrm{imp}}$ does not matter.

\section{Summary of Main Results}
In this section, we briefly summarize the main result obtained via Bethe Ansatz keeping aside the details for later sections. When antiferromagnetic coupling between the bulk $XX$ chain and an impurity is considered, the model has two distinct phases:

\begin{enumerate}
    \item \underline{The Kondo phase} characterized by the screening of the impurity by a multiparticle Kondo cloud exists when the ratio of boundary and bulk coupling is less than the critical value $\frac{J_{\mathrm{imp}}}{J}<\sqrt{2}$. It is often easier to distinguish two subphases in this phase:
    \begin{enumerate}
        \item \underline{The deep Kondo phase} exists when $\frac{J_{\mathrm{imp}}}{J}<1$  where the impurity density of states is Lorentzian-like centered at the Fermi level $E=0$ and the impurity magnetization in the presence of magnetic field crosses over from the screened impurity with $\mathcal{M}_{\mathrm{imp}}=0$ to a free spin $\mathcal{M}_{\mathrm{imp}}=\frac{1}{2}$  at large magnetic field $H_c=2J$ asymptotically just in the case of the conventional Kondo problem \cite{andrei1983solution}.

        \item \underline{The intermediate Kondo phase} exists when $1<\frac{J_{\mathrm{imp}}}{J}<\sqrt{2}$ where the spinons participating in the Kondo cloud form a Lorentzian-like distribution centered at $E=2J$ which is the maximum energy of a single spinon. This shift in the density of the state peak affects other physical quantities such as local impurity magnetization, which again smoothly crosses over from $0$ at $H=0$ to $\frac{1}{2}$ at the critical value of the field $H_c=2J$, but the magnetization curve is a concave upward increasing curve which does not asymptotically reach the maximum value $\frac{1}{2}$ but rather reaches in abrupt manner.
    \end{enumerate} 
   \underline{At the transition point} between the intermediate and deep kondo regime occurs when the boundary coupling is equal to the bulk coupling, at $J_{\mathrm{imp}}=J$. In this case, the impurity density of states is a constant and the Kondo scale $T_K$ as a function of parameter $b = \sec^{-1} \left(\frac{J_{imp}}{J}\right)$ has a point of inflection at $b=0$.
   However, the impurity is still screened by a multiparticle cloud at this point as  is evident from the ground state magnetization, the finite field magnetization and the finite temperature susceptibility calculation presented below.
   
   We will show that the ground state is a sea of a long range singlets in the  entire Kondo phase where impurity is screened by multi-particle Kondo cloud. All excited states are constructed by adding even numbers of spinons on top of the ground state.
   The susceptibility at zero temperature is finite which shows that the impurity is screened at low temperature. However, the susceptibility falls off as $\frac{1}{T}$ at high temperature which indicates that the impurity behaves as a free spin at high temperature.

\item \underline{The bound mode phase} is characterized by the screening of the impurity by a single exponentially localized bound mode with energy $E_b=-2 J J_{\mathrm{imp}}(J_{\mathrm{imp}}^2-J^2)^{-\frac{1}{2}}$ formed at the impurity site when the boundary to the bulk coupling ratio is $\frac{J_{\mathrm{imp}}}{J}>\sqrt{2}$. The bound mode is described by a purely imaginary solution of the Bethe Ansatz equations called \textit{boundary string solution}. This phase is characterized by the impurity density of states given by Eq.\eqref{impdosbm}, which is negative other than a positive delta function contribution from the bound mode. This shows that the screening is effectively a single-particle phenomenon in this regime. Moreover, the local magnetization at the impurity site in the presence of a global magnetic field $H$ has a discontinuity at $H=E_b$ because the massive bound mode screens the impurity. Once the magnetic field has enough energy to flip the bound mode, the impurity magnetization abruptly jumps to $\frac{1}{2}$ as shown in Fig.\ref{fig:pltbm}. Moreover, the susceptibility is finite but negative at zero temperature showing that the impurity is screened and behaves diamagnetically. However, as the temperature increases, the susceptibility becomes positive and eventually falls off as $\frac{1}{T}$ as a free spin. 

Note that in this phase, apart from the usual bulk excitation, a unique boundary excitation is possible. The boundary excitation involves removal of the boundary string and addition of a hole. Impurity is unscreened in such excited states. Hence, the excited states are uniquely sorted in two towers in this phase where one towers includes all the states where impurity is screened and another tower encompasses all the states where impurity is unscreened. This is in sharp contrast with the Kondo phase were there is just a single tower of excited state where impurity is screened in all of the eigenstates.  
\end{enumerate}

 \begin{figure}[H]
    \centering
    \includegraphics[scale=0.75]{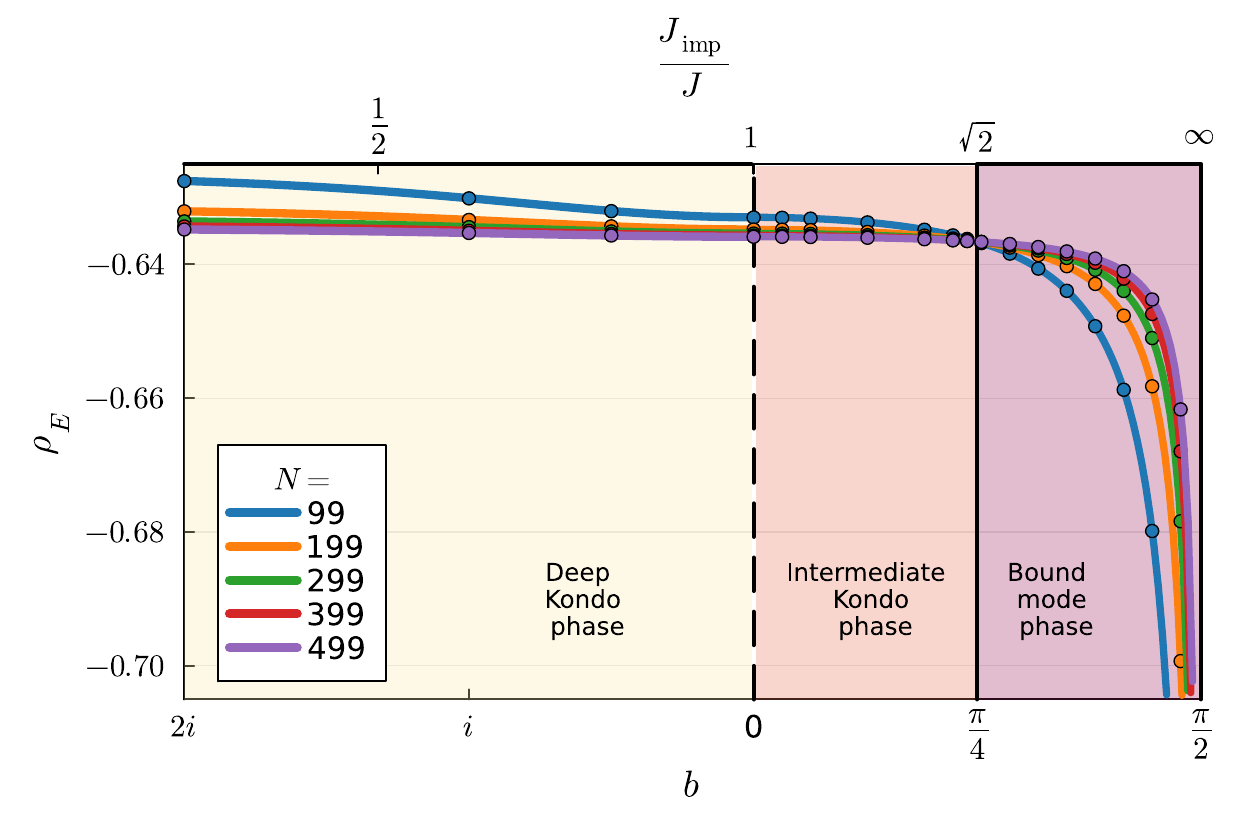}
    \caption{Phase diagram of the model showing the Kondo phase (along with two subphases) and the bound mode phase. The vertical axis represents the ground state energy density $\rho_{E}=\frac{E_{\ket{gs}}}{N+1}$, which is computed using both Bethe Ansatz (solid lines) and the exact diagonalization (circular plot marker) and $x-$axis is the impurity parameter, and both variables  $\frac{J_{\mathrm{imp}}}{J}$ (top) and  $b$, where $\sec(b)=\frac{J_{\mathrm{imp}}}{J}$ (bottom) are shown. Here $N$ is the total number of bulk sites. Notice that at the phase transition point $b=\frac{\pi}{4}$ (or $\frac{J_{\mathrm{imp}}}{J}=\sqrt{2}$), the ground state energy density becomes independent of the system size. The boundary physics depends only on $|J_{\mathrm{imp}}|$. Thus only the case with positive coupling is shown in the diagram.}
    \label{fig:PD}
\end{figure}

\section{The Bethe Ansatz Equations}

To analytically diagonalize the Hamiltonian in  Eq.(\ref{hammodel}), we  proceed to construct the transfer matrix associated with it.  Starting from a spectral parameter $u$ dependent R-matrix $R(u)\in \mathrm{End}(V\otimes V)$ of the tensor square of the vector space $V$ with an explicit matrix form
\begin{equation}
    R(u)=\left(
\begin{array}{cccc}
 a(u) & 0 & 0 & 0 \\
 0 & b(u) & c(u) & 0 \\
 0 & c(u) & b(u) & 0 \\
 0 & 0 & 0 & a(u) \\
\end{array}
\right),
\label{rmat}
\end{equation}
we impose the free fermion condition \cite{maillard1996comment}
\begin{equation}
    a^2(u)+b^2(u)=c^2(u),
\end{equation}
which can be satisfied by choosing
\begin{align}
    a(u)=\cos(u), 
    b(u)=\sin(u), \text{ and } 
    c(u)= 1.
    \label{freefermioncond}
\end{align}

One can readily check that $R(u)$ given by Eq.\eqref{rmat} with entries Eq.\eqref{freefermioncond} is a unitary solution of the the Yang-Baxter equation
\begin{equation}
R_{12}(u-u')R_{13}(u)R_{23}(u')=R_{23}(u')R_{13}(u)R_{12}(u-u'),
\end{equation}
satisfying the unitary condition
\begin{equation}
    R_{12}(u)R_{21}(-u)\propto \mathbf{I}.
\end{equation}

To define the system with open boundary conditions \cite{sklyanin1988boundary}, we introduce the double row monodromy matrix 
\begin{equation}
    \Xi_A(u)=T_A(u)\hat T_A(u),
\end{equation}
where the two single row monodromy matrices are 
 \begin{align}
T_A(u)&=R_{A,N}(u)R_{A,N-1}(u)\cdots R_{A,1}(u)R_{A,0}(u-b)\nonumber\\
\hat T_A(u)&=R_{A,0}(u+b)R_{A,1}(u)\cdots R_{A,N-1}(u)R_{A,N}(u).\nonumber
\end{align}  
and obtain the transfer matrix,
\begin{equation}
    t(u)=\operatorname{tr}_A(\Xi(u)).
\end{equation}
Here, $A$ is the auxiliary space and $0$ is the impurity site and $1$ through $N$ are the labels for the bulk sites. Note that the inhomogeneity parameter $b$ at the $0^{th}$ site located left of site 1 will give the impurity term in the Hamiltonian via $J_{\mathrm{imp}}=J\sec(b)$ as shown in Eq.\eqref{hamrel}. 

The Hamiltonian is then obtained from the transfer matrix as
\begin{align}
    H&=J\frac{\mathrm{d}}{\mathrm{d}u}\log t(u)\Big\vert_{u\to 0}\nonumber\\
    &=\frac{1}{2} \left[\sum_{j=1}^{N-1} J (\sigma_j^x \sigma_{j+1}^x + \sigma_j^y \sigma_{j+1}^y) +J\sec(b) (\sigma_1^x \sigma_{0}^x + \sigma_1^y \sigma_{0}^y)\right].
    \label{hamrel}
\end{align}
This Hamiltonian acts on the product space
$\otimes_{j=1}^N \frak{h}_j$ where the local Hilbert space of each quantum spin variable is the two-dimensional complex vector space $\frak{h}_n=\mathbb{C}^2$ and $\sigma^x_j$ and $\sigma^y_j$ are the Pauli operators acting in $\frak{h}_j$. Notice that the impurity coupling $J_{\mathrm{imp}}=J\sec(b)$, is parameterized by the variable $b$, so that for purely imaginary $b$, the boundary coupling $J_{\mathrm{imp}}<J$ and for $0<b<\frac{\pi}{2}$, it is between $J$ and infinity. This parameterization is natural in the Bethe Ansatz and often makes the resulting expressions simpler.  Notice that the spectrum of the model is the same for $\pm J_{\mathrm{imp}}$; thus only positive $J_{\mathrm{imp}}$ is considered in the above parameterization.

We employ the functional Bethe Ansatz method \cite{sklyanin1990functional} to diagonalize the transfer matrix and obtain the Bethe Ansatz equations. We recall that the quantum determinants of the single row monodromy matrices are given as\cite{kulish1992algebraic}

\begin{align}
& \operatorname{Det}_q\{T(u)\}=\operatorname{tr}_{1,2}\left\{P_{1,2}^{(-)} T_1\left(u-\frac{\pi}{2}\right) T_2(u) P_{1,2}^{(-)}\right\} \\
& \operatorname{Det}_q\{\hat{T}(u)\}=\operatorname{tr}_{1,2}\left\{P_{1,2}^{(-)} \hat{T}_1\left(u-\frac{\pi}{2}\right) \hat{T}_2(u) P_{1,2}^{(-)}\right\},
\end{align}

where
\begin{equation}
    P_{1,2}^{(-)}=\frac{1-R(0)}{2}=-\frac{1}{2}R\left(-\frac{\pi}{2} \right)
\end{equation}
is the antisymmetric projection operator $\left(P_{1,2}^{(-)}\right)^2=P_{1,2}^{(-)}$.

We can compute explicitly
$$
\operatorname{Det}_q(R(u))=\operatorname{tr}_{1,2}\left\{P_{1,2}^{(-)} R_{1, j}\left(u-\frac{\pi}{2}\right) R_{2, j}(u) P_{1,2}^{(-)}\right\}=\sin \left(u-\frac{\pi}{2}\right) \sin \left(u+\frac{\pi}{2}\right)
$$
and obtain the quantum determinant of the transfer matrix via the relation
\begin{equation}
    \operatorname{Det}_q\{T(u)\}=\prod_{j=1}^N \operatorname{Det}_q\{R(u)\}.
\end{equation}

The eigenvalue of the transfer matrix satisfy the relation \cite{wang2015off}
\begin{equation}
    \begin{aligned}
\Lambda\left(\theta_j\right) \Lambda\left(\theta_j-\frac{\pi}{2}\right) & =\frac{\Delta_q\left(\theta_j\right)}{\cos ^2\left(2 \theta _j\right)} =a\left(\theta_j\right) d\left(\theta_j-\frac{\pi}{2}\right), \quad j=1, \ldots, N,
\end{aligned}
\end{equation}
where
\begin{equation}
    \Delta_q(u)=\operatorname{Det}_q\{T(u)\} \operatorname{Det}_q\{\hat{T}(u)\} 
\end{equation}

such that
\begin{align}
& a(u)=2 \frac{\cos ^2(u)}{\cos(2u)} {\cos (u-b) \cos (u+b)} {\cos ^{2 N}(u)} \\
& d(u)=a\left(-u-\frac{\pi}{2}\right)=2\frac{\sin(u)^2}{\cos(2u)}{\sin (u-b) \sin (u+b)} {\sin ^{2 N}(u)}.
\end{align}
Baxter's T-Q relation for the eigenvalue can be written as \cite{baxter2016exactly}
\begin{equation}
    \Lambda(u)=a(u) \frac{{Q}(u-\frac{\pi}{2})}{{Q}(u)}+d(u) \frac{{Q}(u+\frac{\pi}{2})}{{Q}(u)},
    \label{lambdaeqnn}
\end{equation}
where the $ Q$ function is
\begin{equation}
  {Q}(u)=\prod_{\ell=1}^M {\sin \left(u-u_{\ell}\right) \cos \left(u+u_{\ell}\right)}.
\end{equation}

Regularity of the T–Q equation gives the Bethe Ansatz equations
\begin{equation}
    \begin{gathered}
\left(\frac{\cos \left(u_j\right)}{\sin \left(u_j\right)}\right)^{2 N+2} \frac{\cos \left(u_j+b\right)}{\sin \left(u_j+b\right)} \frac{\cos \left(u_j-b\right)}{\sin \left(u_j-b\right)} =1.
\end{gathered}
\end{equation}
Upon making the transformation $u_j=i\frac{\lambda_j}{2}-\frac{\pi}{4}$, the Bethe Ansatz equations become
\begin{equation}
    \left(\frac{\sinh \left(\frac{\lambda_j}{2}+\frac{i \pi}{4} \right)}{\sinh \left(\frac{\lambda_j}{2}-\frac{i \pi}{4} \right)}\right)^{2 N+2}\frac{\sinh \left(\frac{\lambda_j}{2}+i  b+\frac{i \pi}{4} \right)}{\sinh \left(\frac{\lambda_j}{2}-i b-\frac{i \pi}{4} \right)}  \frac{\sinh \left(\frac{\lambda_j}{2}-i  b+\frac{i \pi}{4} \right)}{\sinh \left(\frac{\lambda_j}{2}+i b-\frac{i \pi}{4} \right)}   =1.
    \label{baebae}
\end{equation}
As shown in \ref{wfnsec}, the Bethe Ansatz equation \eqref{baebae} is the quantization condition written for the rapidity $\lambda$ related to the quasimomenta $k$ via $\frac{\sinh \left(\frac{\lambda_j}{2}+\frac{i \pi}{4} \right)}{\sinh \left(\frac{\lambda_j}{2}-\frac{i \pi}{4} \right)}=e^{-ik_j}$. Because the quantization condition for $k$ is a transcendental equation, writing it as Bethe Ansatz equations $\eqref{baebae}$ makes it easier to analytically study the model.

 The energy eigenvalues follow from Eq.\eqref{hamrel} and Eq.\eqref{lambdaeqnn}
\begin{equation}
    E=J\left.\frac{\mathrm{d}}{\mathrm{d}\lambda}\Lambda(\lambda)\right|_{\lambda\to 0}=-2J \sum_j \frac{1}{\cosh(\lambda_j)}
    \label{engrel}.
\end{equation}

\section{Results}

The solutions of the Bethe Ansatz equations Eq.\eqref{baebae} depend on the value of the parameter $b$.  In addition to the standard (real) solutions that describe the dynamics of spin flips on the chain, the Bethe Ansatz equation Eq.\eqref{baebae} has a unique purely imaginary solution in the thermodynamic limit of the form
\begin{equation}
    \lambda_b=\frac{i}{2}\left(4b-\pi\right),
\end{equation}
when $\frac{\pi}{4}<b<\frac{\pi}{2}$, corresponding coupling strength $<\sqrt{2}J<J_{\mathrm{imp}}$; while when either $b$ is purely imaginary or it takes real values in the range $0<b<\frac{\pi}{4}$, there is no such solution. This solution is called the boundary string solution. Such a solution describing a bound mode appears in many one-dimensional models with boundaries \cite{kapustin1996surface,pasnoori2020kondo,pasnoori2021boundary,pasnoori2022rise,kattel2023exact,rylands2020exact,wang1997exact,frahm1997open}.

We will now solve the Bethe equations in these two regimes separately. We shall refer to the phase in which no boundary string solution exists as the  ``Kondo phase", while the phase in which the boundary string solution exists will be termed the  ``Bound mode phase", and the rationale for this nomenclature will become evident shortly. In both phases, we describe the ground state, the bulk and boundary elementary excitations, the effects of magnetic field and the finite temperature effects.

{\color{blue}\subsection{Kondo Phase}}

 The Kondo phase corresponds to the range where 
  $0<J_{\mathrm{imp}}<\sqrt{2}J$, or the parameter $b$ being purely imaginary or between $0<b<\frac{\pi}{4}$. We call the 
  the parameter regime where the parameter $b$ is imaginary ($0<J_{\mathrm{imp}} <J$), the deep Kondo regime and the parameter range $0<b<\frac{\pi}{4}$ ($J<J_{\mathrm{imp}} < \sqrt2 J$) the intermediate Kondo regime.\\

Taking $\log$ on both sides of Eq.(\ref{baebae}) and differentiating with respect to $\lambda$ leads to
\begin{equation}
2\rho(\lambda)=\frac{\text{sech}(\lambda )}{\pi } \left(2 \cosh ^2(\lambda ) \left(\frac{\cos (2 b)}{\cos (4 b)+\cosh (2 \lambda )}\right)+N+1\right)-\delta(\lambda).
\label{rootdens}
\end{equation}
The delta function is added to remove the root at the origin because the trivial solution $\lambda=0$ ($k=0$) leads to a non-normalizable vanishing wavefunction. 
Note that the root distribution naturally separates into the bulk, the boundary and the impurity part as 
\begin{equation}
    \rho(\lambda) = \rho_{\mathrm{bulk}}N+\rho_{\mathrm{imp}}+\rho_\mathrm{boundary},
\end{equation}
where the impurity contribution is
\begin{equation}
    \rho_\mathrm{imp} = \frac{2}{\pi }  \cosh (\lambda ) \left(\frac{\cos (2 b)}{\cos (4 b)+\cosh (2 \lambda )}\right).
\end{equation}
As usual, for an odd number of total sites, a zero-energy spinon has to be added in the ground state. 

The ground state energy for odd $N$ is given by
\begin{equation}
    E_{gs}=-2J\int_{-\infty}^\infty \frac{1}{\cosh(\lambda)}\rho(\lambda)\mathrm{d}\lambda=-\frac{2 J (N+1)}{\pi }-\frac{4 b J \csc (2 b)}{\pi }+J.
    \label{gsK}
\end{equation}

All other excitations are constructed by adding even number of holes with energy
\begin{equation}
    E_\theta=\frac{2J}{\cosh(\theta)},
\end{equation}
where $\theta$ is the position of the hole. When $\theta\to \pm \infty$, $E_\theta\to 0$ which shows that the model is gapless. There are no boundary excitations in this phase.

From the root density Eq.\eqref{rootdens}, we obtain the ratio of the density of states contribution by the impurity to the bulk as
\begin{equation}
R(E)=\frac{N}{2} \frac{\rho_{\text {dos }}^{\text {imp }}(E)}{\rho_{\text {dos }}^{\text {bulk }}(E)}=\frac{4 J^2 \cos (2 b)}{E^2\left(\cos (4 b)+\cosh \left(2 \cosh ^{-1}\left(\frac{2 J}{E}\right)\right)\right)},
\label{redef}
\end{equation}
where we used
\begin{equation}
    E_\theta=2J\text{sech}(\theta),
\end{equation}
the energy of a single spinon which can range from $0$ to its maximum $2J$ to invert the relation.
\begin{figure}[H]
    \centering
    \includegraphics[scale=0.45]{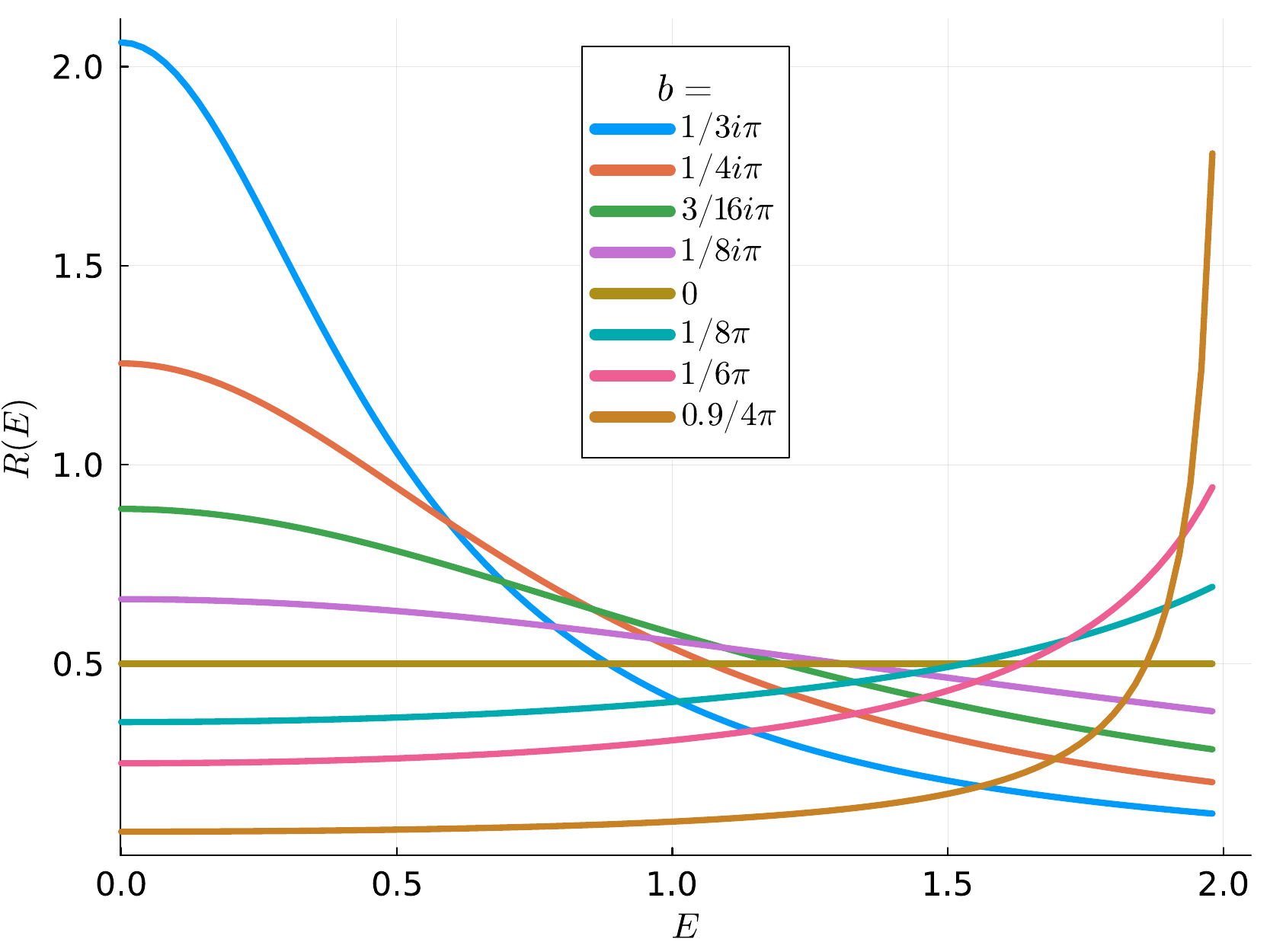}
  \caption{The plot presents the spectral weight of the spinons in the Kondo cloud screening impurity $R(E)$given in Eq.\eqref{redef}. This spectral weight was computed using the \textit{Bethe Ansatz} method. The plot reveals intriguing trends: in the deep Kondo regime, the majority of spinons involved in impurity screening have energies close to the Fermi energy ($E=0$). However, as the impurity coupling surpasses that of the bulk such that the model is in intermediate Kondo phase most of the screening spinons clustering around an energy level close to $2J$, which is the maximum energy of a single spinon.}
\end{figure}
Only in the deep Kondo regime \textit{i.e.}when $0<J_{\mathrm{imp}}<J$, the ratio of the impurity to the bulk density of states $R(E)$ takes a characteristic Kondo Lorentzian-like peak centered at $E=0$. However in the intermediate Kondo regime \textit{i.e.} when $J< J_{\mathrm{imp}}< \sqrt{2}J$, the peak shifts from the Fermi surface $E=0$ to the maximum energy of the spinon $2 J$. At $b=0$, the impurity coupling is the same as the bulk coupling. Thus, the impurity becomes the part of the bulk and at this point, $R(E)$ becomes a constant function. This change in the shape of magnetization curve and the peak of Lorentzian from $E=0$ to $E=2J$ in the intermediate Kondo phase shows that the model is preparing an announcement of the bound mode phase for $b>\frac{\pi}{4}$ where the impurity is screened by a single mode thus characterized by a delta function peak in the density of state as show later in Eq.\eqref{impdosbm}.

Given the density of states,  we define the Kondo temperature as the energy scale at which the integrated impurity density of the state is half of the total number of state contributed by impurity \textit{i.e.}
\begin{equation}
    \int_0^{T_K} \mathrm{d} E \rho_{\text {dos }}^{\text {imp }}(E)=\frac{1}{2} \int_0^{2J} \mathrm{d} E \rho_{\text {dos }}^{\text {imp }}(E),
\end{equation}
where the integral bound of over the possible energy window of a single spinon. We thus obtain,
\begin{equation}
    T_K=\frac{2 J \sqrt{1-\cos (2 b)}}{\sqrt{1-\cos ^2(2 b)}}=\sqrt{2}J\sec(b).
    \label{tkdefdef}
\end{equation}
\begin{figure}[H]
    \centering
    \includegraphics[scale=0.5]{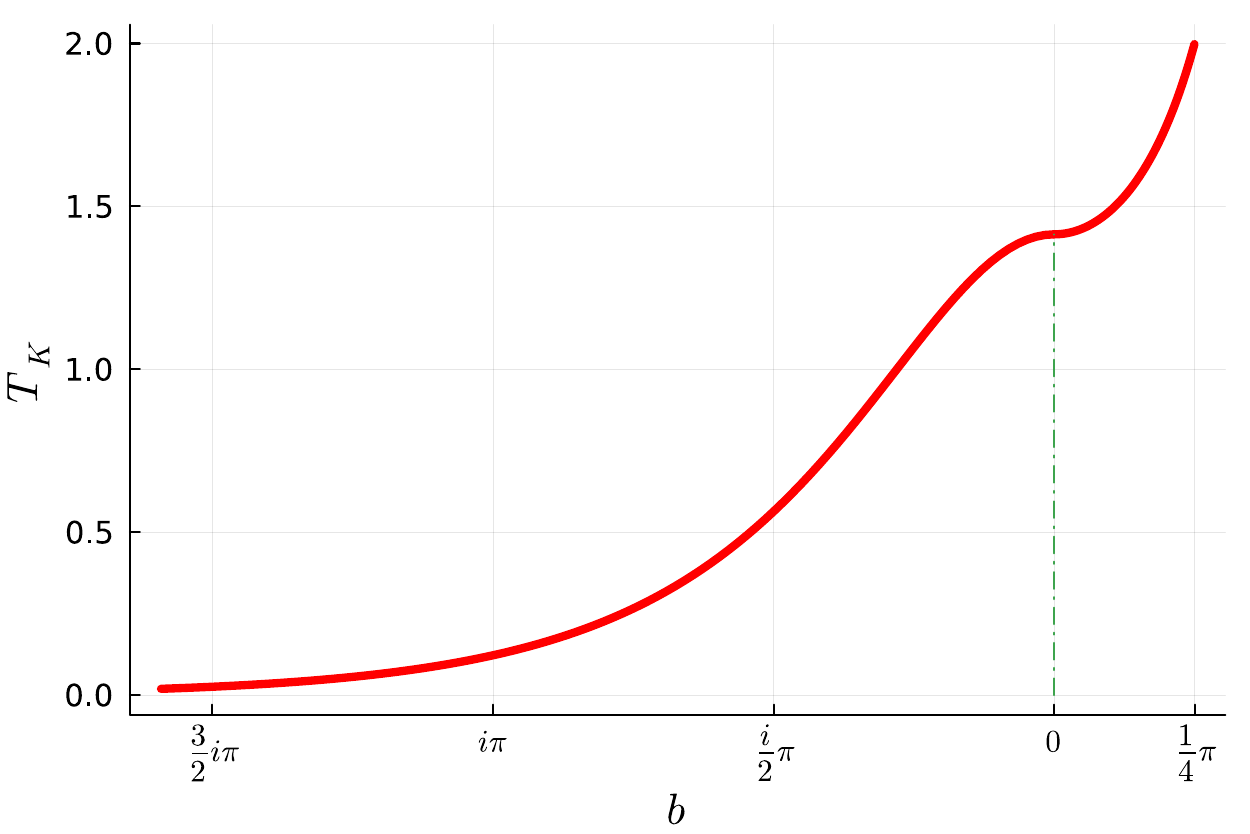}
    \caption{The characteristic Kondo scale within the Kondo regime. The vertical line separates the deep and intermediate Kondo phase. At the phase transition point shown by vertical dashed line, the scale $T_K$ has an inflection point.}
    \label{fig:TK-Kondo}
\end{figure}
As shown the the figure Fig.\ref{fig:TK-Kondo}, the Kondo temperature increases as imaginary $b$ is decreases (or equivalently $J_{\mathrm{imp}}$ increases). When $b=0$, $J_{\mathrm{imp}}=J$ and $T_K=\sqrt{2}J$. Finally when $b$ reaches the critical value $\frac{\pi}{4}$, $J_{\mathrm{imp}}=\sqrt{2}J$, then $T_K=2J$, which is the maximum energy of possible for a single spinon. Notice the inflection point in the plot of $T_K$ in Fig.\ref{fig:TK-Kondo} at $b=0$. As mentioned earlier, at $b=0$, the boundary and bulk couplings are the same where the ratio of the impurity to bulk density of state becomes a constant. This point demarcates the deep and intermediate kondo phase.

\subsubsection{\color{blue}{Effect of magnetic field}}\phantom{}\\

We now consider the system in the presence of a magnetic field H and add the magnetic term
\begin{equation}
  \mathcal{H}_{\mathrm{mag}}=  -HS_L^z-H\sum_{i=1}^N S_i^z.
\end{equation}
In the fermionic language this term corresponds to applying a chemical potential.  It commutes with the Hamiltonian given by Eq.\eqref{hammodel}, so the eigenstate of $\mathcal{H}$ also diagonalizes the Hamiltonian $\mathcal{H}_{T}=\mathcal{H}+\mathcal{H}_{\mathrm{mag}}$. However, the ground state will change as spin starts to flip to allign in the direction of the applied field. Magnetization of the impurity can be computed by minimizing the energy in the presence of the magnetic field, which is given by

\begin{equation}
    E_B=-2J\int_{-B}^B \frac{1}{\cosh(\lambda)}\rho_B(\lambda)\mathrm{d}\lambda-\mathcal{M} H
    \label{engrell}
\end{equation}
where
\begin{equation}
    \mathcal{M}=\frac{N+1}{2}-\int_{-B}^B \rho_B(\lambda)\mathrm{d}\lambda
    \label{magmageqnexp}
\end{equation}
is the magnatization and $B<\infty$ is the yet undetermined new Fermi level in the presence of external magnetic field. Here the root density $\rho_B(\lambda)$ in the presence of the magnetic field is same as $\rho(\lambda)$ because of the lack of backflow effect. 
Upon minimizing the energy Eq.\eqref{engrell}, we obtain the relation between the integral bound $B$ and the magnetic field $H$ as
\begin{equation}
    B=\cosh ^{-1}\left(\frac{2 J}{H}\right).
\end{equation}
From Eq.\eqref{magmageqnexp}, the magnetization solely due to impurity becomes 
\begin{equation}
    \mathcal{M}_{\mathrm{imp}}=\frac{1}{2}-\int_{-B}^B \rho_\mathrm{imp}(\lambda)\mathrm{d}\lambda=\frac{1}{2}-\frac{\tan ^{-1}\left(\sec (2 b) \sqrt{\left(\frac{2 J}{H}-1\right) \left(\frac{2 J}{H}+1\right)}\right)}{\pi }.
    \label{magimp}
\end{equation}
Notice that
\begin{equation}
    \underset{H\to 0^+}{\text{lim}}\frac{\tan ^{-1}\left(\sec (2 b) \sqrt{\left(\frac{2 J}{H}-1\right) \left(\frac{2 J}{H}+1\right)}\right)}{\pi } =\frac{1}{2},
\end{equation}
    
which shows that the impurity spin is completely quenched in the ground state by the Kondo cloud. 

\begin{figure}[H]
    \centering
    \includegraphics[scale=0.5]{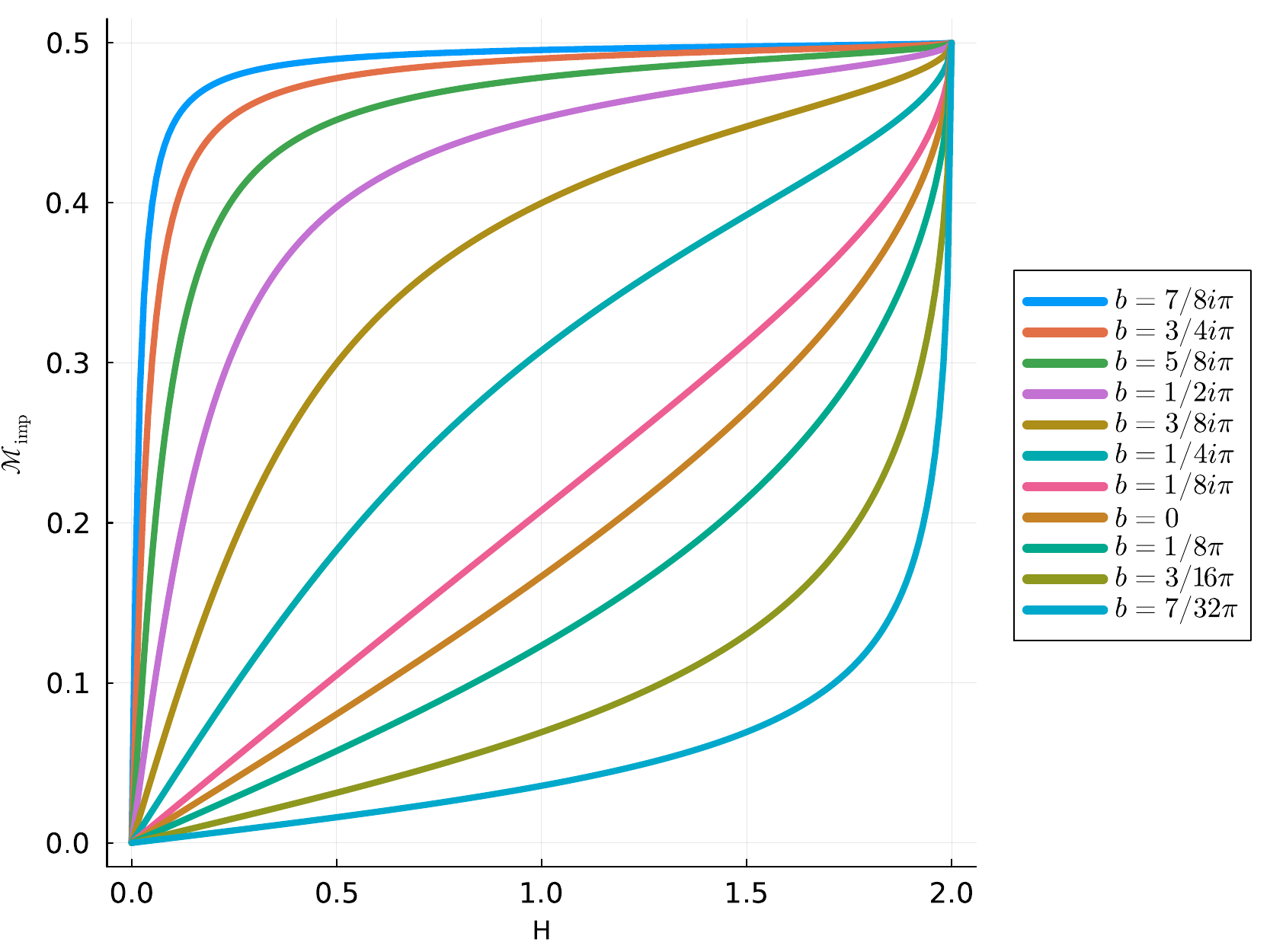}
    \caption{Impurity magnetization in the Kondo phase given by Eq.\eqref{magimp} obtained from Bethe Ansatz shows that the impurity spin is screened at low field and it behaves as a free impurity at high field $H>2J$ just like in conventional Kondo problem. Notice that the shape of the curves are different for imaginary $b$ (deep Kondo phase) and real $b$ (intermediate Kondo phase).}
    \label{fig:pltkp}
\end{figure}

As shown in Fig.\ref{fig:pltkp}, the behavior of the impurity magnetization curve is different in the deep and intermediate Kondo regimes. In the deep Kondo regime where the impurity parameter $b$ is imaginary, the impurity magnetization is a concave function where the magnetization smoothly increases from $0$ at $H=0$ and reaches $\frac{1}{2}$ at the critical value $H_c=2J$ asymptotically. However, in the intermediate Kondo phase when $0<b<\frac{\pi}{4}$, the magnetization is a convex function which grows from $0$ at $H=0$ to $\frac{1}{2}$ at the critical value $H_c=2J$ following a concave upward trajectory. 

For each of the bulk spins, the magnetization is given by
\begin{equation}
    \mathcal{M}_{\text{bulk}}=\frac{1}{2}-\frac{\tan ^{-1}\left(\sqrt{\left(\frac{2 J}{H}-1\right) \left(\frac{2 J}{H}+1\right)}\right)}{\pi }
\end{equation}

Thus, in this phase, there is a critical magnetic field $H_c=2J$ at which each spin including the impurity spin is fully polarized. 


 From Eq.\eqref{tkdefdef}, we obtain
\begin{equation}
  \cos(2b)=\frac{4 J^2-T_{\text{K}}^2}{T_{\text{K}}^2}.
  \label{cs2b}
\end{equation}

such that the impurity magnetization becomes
\begin{equation}
    \mathcal{M}_{\mathrm{imp}}=\frac{1}{2}-\frac{\tan ^{-1}\left(\frac{\text{Tk}^2 \sqrt{\frac{2 J}{H}-1} \sqrt{\frac{2 J}{H}+1}}{4 J^2-\text{Tk}^2}\right)}{\pi }.
    \label{impmagrel}
    \end{equation}

For deep Kondo regime which is when $b$ is purely imaginary, expanding the impurity magnetization around the critical value of magnetic field $H_c=2J$, we obtain the asymptotic
\begin{equation}
    \mathcal{M}_{\mathrm{imp}} (H\to 2J)=\frac{1}{2}-\frac{\sqrt{2-\frac{H}{J}} T_{\text{K}}^2}{\pi  \left(4 J^2-T_{\text{K}}^2\right)}.
    \label{asympmag}
\end{equation}

This result indicates a deviation from the conventional Kondo problem, where the magnetization of the impurities converges to $\frac{1}{2}$ while exhibiting logarithmic corrections \cite{andrei1983solution}. On the contrary, the model examined in this study displays a Kondo phase characterized by impurity magnetization approaching $\frac{1}{2}$ with a power law as in Eq.\eqref{asympmag}.

Note that the impurity magnetization given by Eq.\eqref{magimp} is the magnetization of the impurity computed in the rapidity that includes all the contributions only from the impurity part. For a discrete lattice problem like the spin chain under consideration, we can also ask questions like what is the magnetization at each site. We will now compute the magnetization at the impurity site $\langle S^z(0)\rangle$ in the presence of the global magnetic field $H$. 

To compute this quantity, let us change the variables from rapidity $\lambda_j$ to the quasi momentum $k_j$ using
\begin{equation}    \frac{\sinh\left(\frac{\lambda_j}{2}+\frac{i\pi}{4}\right)}{\sinh\left(\frac{\lambda_j}{2}-\frac{i\pi}{4}\right)}=-e^{-ik_j}
\label{varchangelambdatok}
\end{equation}
 such that the Bethe Ansatz equation Eq.\eqref{baebae} becomes the quantization condition on the quasi momenta
 \begin{equation}
     2 \cos \left(k_j\right) \sin \left((N+1) k_j\right)-\sec ^2(b) \sin \left(N k_j\right)=0,
     \label{kquant}
 \end{equation}
 and the energy relation Eq.\eqref{engrel} becomes
 \begin{equation}
     E=-2J\sum_{k_j}\cos(k_j).
 \end{equation}
 The fermi points are located at $k_F=\pm \frac{\pi}{2}$. 

In the presence of the global magnetic $H$, the Fermi sea is shifted, and now the new Fermi points are located at 
\begin{equation}
    k_F(H)=\pm\cos ^{-1}\left(-\frac{H}{2}\right).
\end{equation}
Thus, the magnetization at the first site is given by
\begin{equation}
    \langle S^z(0)\rangle=\sum_{k<k_F(H)}S_{k}^z(0),
    \label{magimpsite}
\end{equation}
where
\begin{equation}
   S_{k}^z(0)= F_k^*(0)F_k(0)-\frac{1}{2}.
\end{equation}
Here $F(0)$ is the normalized wavefunction at the impurity site. As shown in \ref{wfnsec}, the normalized wavefunction at the impurity site is
\begin{equation}
 F_k(0)=   \frac{\sin (k N)}{\sqrt{\frac{1}{4} \sec ^2(b) (\csc (k) \sin (k-2 k N)+2 N-1)+\sin ^2(k N)}}.
\end{equation}
\begin{figure}[H]
    \centering
    \includegraphics[scale=0.45]{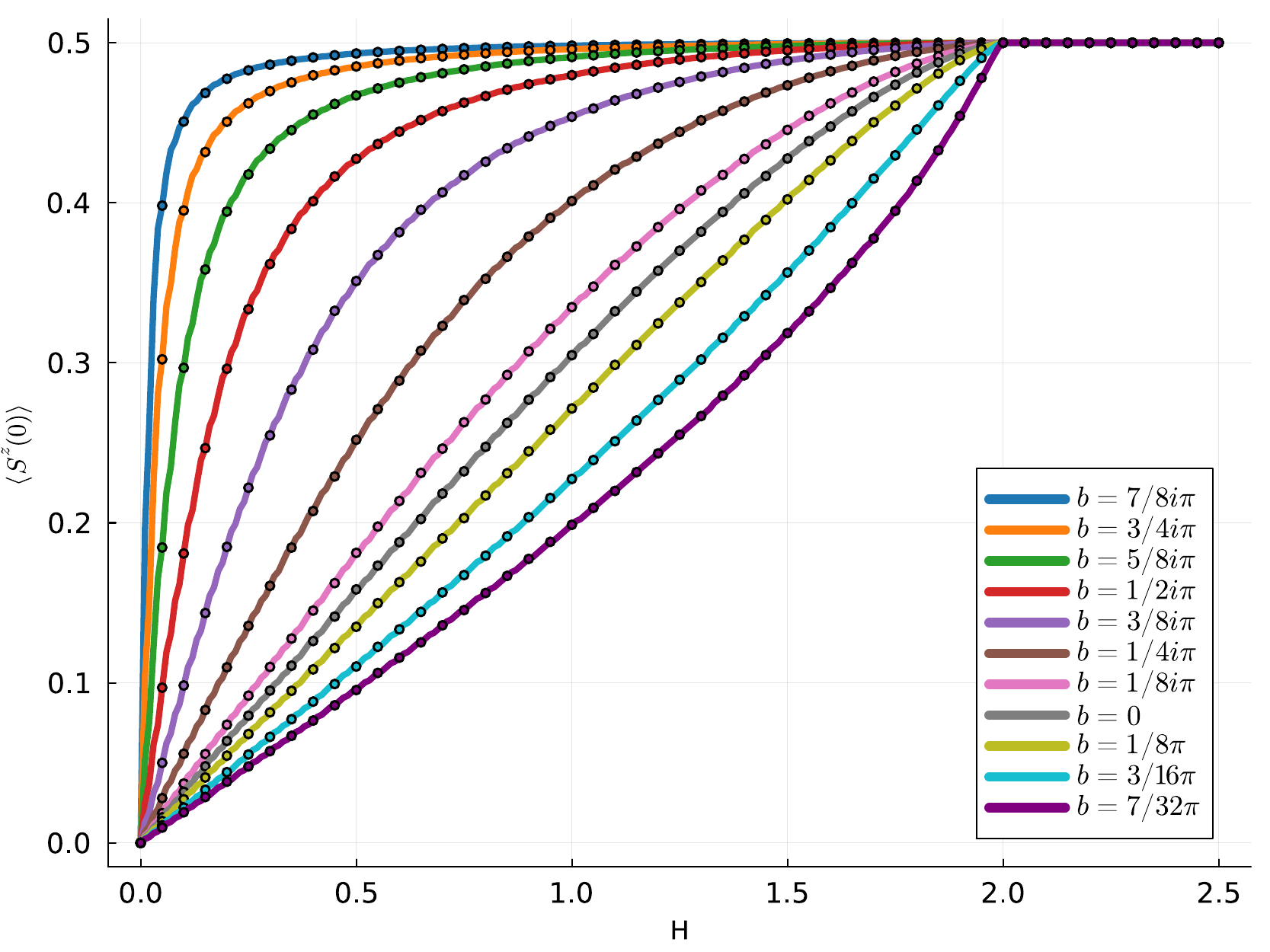}
    \caption{Local magnetization at the impurity site in the Kondo phase. The magnetization grows smoothly and reaches $\frac{1}{2}$ at $H_c=2J$ for all values of $H$ in the Kondo phase for $N=999$. The solid curve is obtained from Eq.\eqref{magimpsite} and the dotted points are obtained from exact diagonalization.}
    \label{fig:impmagK}
\end{figure}

Since the quantization condition, Eq.\eqref{kquant} is a transcendental equation which cannot be solved in closed form. We graphically solve for $k_j$ when $N=999$ (such that the total sites including the impurity is 1000), and plot the magnetization curve for various values of $b<\frac{\pi}{4}$. We also directly calculated the magnetization at the impurity site using exact diagonalization with 999 bulk sites and 1 impurity and show both the curve from Eq.\eqref{magimpsite} and exact diagonalization as shown in Fig.\ref{fig:impmagK}.

\subsubsection{\color{blue}{{Finite temperature effects}}}\phantom{}\\

We now turn to the finite-temperature susceptibility calculation. The free energy is given by
\begin{equation}
    f=-T\int_{0}^\pi \mathrm{d}k~\rho(k)\ln \left(1+ e^{-\frac{2h-2J\cos(k)}{T}} \right).
\end{equation}
Using Eq.\eqref{varchangelambdatok} in the density of the solution of the Bethe equation Eq.\eqref{rootdens} and recalling that the model has particle-hole  symmetry, we can find the density of the quasi-momenta in the thermodynamic limit
\begin{equation}
   \rho(k) = \frac{N+1}{\pi }-\frac{ \cos (2 b)}{\pi  \left(\sin ^2(2 b) \cos ^2(k)-1\right)}-\delta(k).
\end{equation}

Taking the impurity contribution upon using Eq.\eqref{cs2b} becomes
\begin{equation}
    \rho_{\mathrm{imp}}(k)=\frac{4 J^2 T_{\text{K}}^2-T_{\text{K}}^4}{\pi  \left(T_{\text{K}}^4-8 J^2 T_{\text{K}}^2 \cos ^2(k)+16 J^4 \cos ^2(k)\right)}.
\end{equation}

 Thus, the free energy contribution due to the impurity becomes
\begin{equation}
    f_{\mathrm{imp}}=-T\int_0^\pi\mathrm{d}k \frac{4 J^2 T_{\text{K}}^2-T_{\text{K}}^4}{\pi  \left(T_{\text{K}}^4-8 J^2 T_{\text{K}}^2 \cos ^2(k)+16 J^4 \cos ^2(k)\right)}\ln \left(1+ e^{-\frac{2h-2J\cos(k)}{T}} \right),
\end{equation}

such that the finite temperature susceptibility becomes
\begin{equation}
    \chi_{\mathrm{imp}}=-\frac{d^2}{dh^2}f_{\mathrm{imp}} \big\vert_{h\to 0}=\int_0^\pi \frac{ \cos (2 b) \text{sech}^2\left(\frac{J\cos (k)}{T}\right)}{\pi  T \left(1-\sin ^2(2 b) \cos ^2(k)\right)}\mathrm{d}k.
    \label{susint}
\end{equation}

\begin{figure}[H]
    \centering
    \includegraphics[scale=0.65]{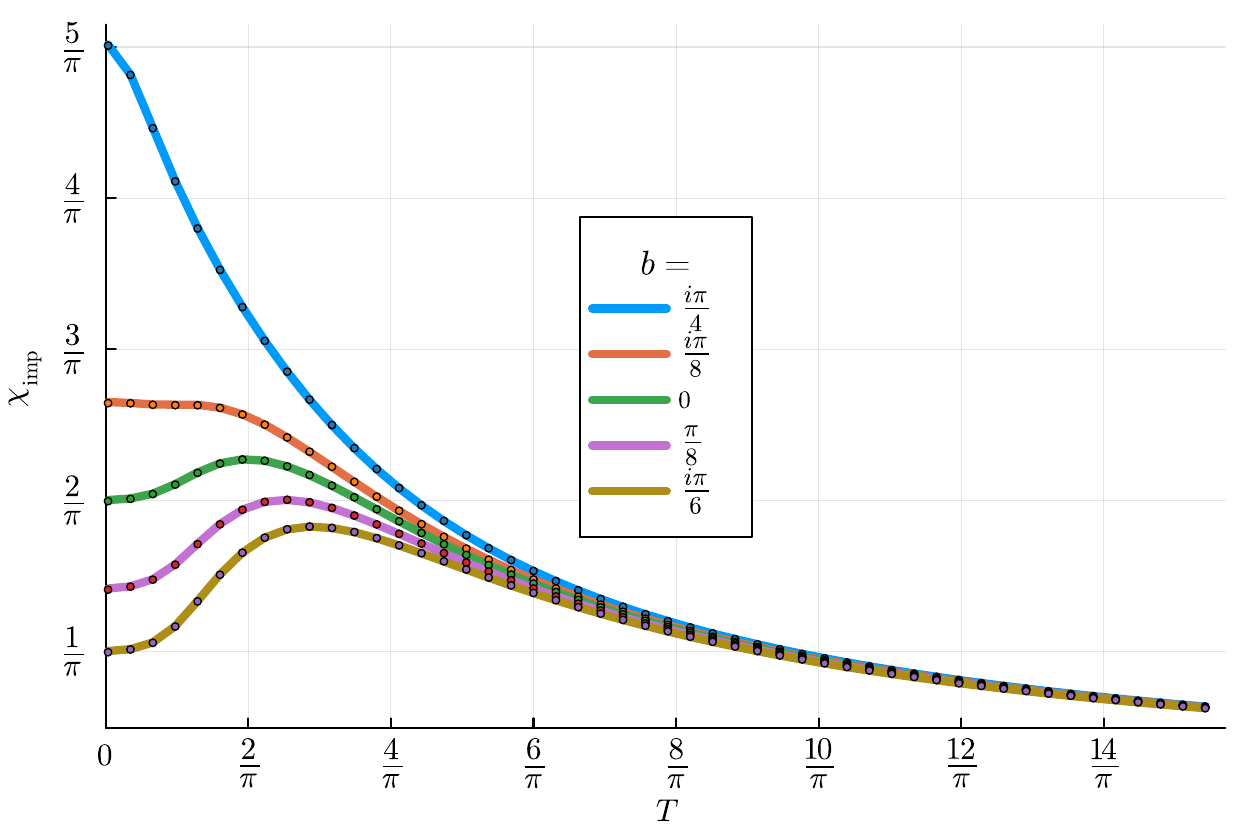}
    \caption{Susceptibility at low temperature for various values of impurity parameter in the Kondo regime. The solid line is obtained by performing numerical integration of Eq.\eqref{susint} and the discrete points are obtained by numerically computing the impurity susceptibility for a chain of total 500 number of sites and $J=1$ using exact diagonalization by subtracting the susceptibility of the chain with 500 site and impurity from the susceptibility of a chain with 499 sites without any impurity.}
    \label{fig:sus}
\end{figure}

The low temperature asymptotic expansion of the susceptibility at 
\begin{equation}
\begin{split}   
    {\chi_{\mathrm{imp}}}_{(T\to 0)}&=\frac{2 \cos (2 b) \tanh \left(\frac{J}{T}\right)}{\pi  J}+\mathcal{O}(T)\\&=\frac{8 J \tanh \left(\frac{J}{T}\right)}{\pi  T_{\text{K}}^2}-\frac{2 \tanh \left(\frac{J}{T}\right)}{\pi  J}+\mathcal{O}(T)
\end{split}
\end{equation}
The susceptibility at $T=0$ is
\begin{equation}
    \chi_\mathrm{imp}(T=0)=\frac{2}{\pi J}\cos(2b)= \frac{8J}{\pi T_k^2}-\frac{2}{\pi J}.
\end{equation}
This finite value of susceptibility at $T=0$ shows that the impurity is screened at low temperature. Now, looking at the asymptotic of the integrand at $T\to \infty$, we obtain
\begin{align}
{\chi_{\mathrm{imp}}}_{(T\to\infty)}&=\int_0^\pi \frac{4 \cos (2 b)}{\pi  T \left(-2 \sin ^2(2 b) \cos (2 k)+\cos (4 b)+3\right)}\mathrm{d}k\nonumber\\
&=\frac{1}{T}\left(1-\frac{T_{\text{K}}^2}{4 T^2}\frac{+J^2 T_{\text{K}}^2}{12 T^4}+\frac{T_{\text{K}}^4}{48 T^4}\right)+\mathcal{O}\left(\frac{1}{T^6}\right)
\end{align}
when $0<b<\frac{\pi}{4}$ or $b$ is purely imaginary, \textit{i.e.} in the entire Kondo regime. In the last step, we used Eq.\eqref{cs2b}. This Curie-like susceptibility as high temperature shows that the impurity behaves like a free spin at high temperature.

Before we proceed to discuss the physics in the bound mode regime, we would like to remind that the impurity is screened by multiparticle Kondo cloud in all eigenstates of the model in this phase at zero temperature and zero field. We will see that, in the bound mode phase, there are two distinct kinds of eigenstates: one where the impurity in screened by a localized  bound mode and one where where impurity is unscreened. 

{\color{blue}\subsection{Bound mode Phase}}
We now solve the Bethe Ansatz equation in the parametric regimes $\frac{\pi}{4}<b<\frac{\pi}{2}$. Notice that in this regime, the Bethe Ansatz equation Eq.\eqref{baebae} has a unique purely imaginary solution of the form
\begin{equation}
    \lambda_b=\frac{i}{2}\left(4b-\pi\right)
\end{equation}
on top of other real $\lambda$ solutions. This solution is called the boundary string solution and describes the boundary bound mode that exists at the boundary of various one-dimensional integrable models \cite{kapustin1996surface,leclair1995boundary,skorik1995boundary,tsuchiya1997boundary,rylands2020exact,pasnoori2021boundary,pasnoori2023boundary,kattel2023kondo}.

The energy of this solution is
\begin{equation}
    E_b=-2 J \csc (2 b).
    \label{bstringeng}
\end{equation}
The energy of the boundary string ranges from $-2J<E_b<-\infty$ as shown in the figure below. Since, the boundary string has negative energy, it exists in the ground state when $\frac{\pi}{4}<b<\frac{\pi}{2}$. The ground state is made up of all real root described by the continuous root distribution $\rho(\lambda)$ and the isolated purely imaginary boundary string solution. The impurity is screened by the bound mode formed at the impurity site. A unique excited state with boundary excitations can be constructed by removing the boundary string solution from the ground state and adding a hole. The state thus constructed contains an unscreened impurity. 
\begin{figure}[H]
\centering
\includegraphics[scale=0.4]{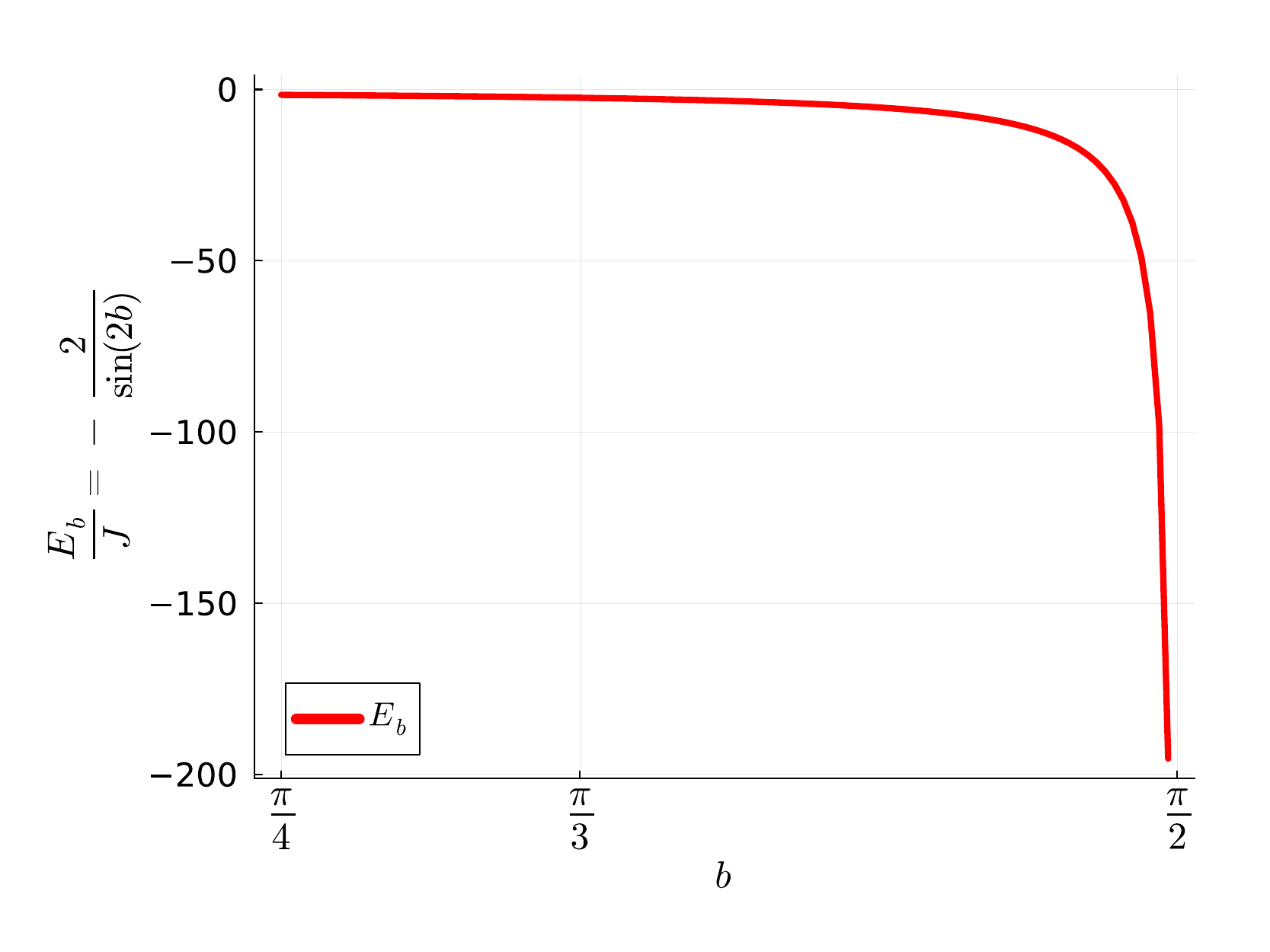}
\caption{Boundary string energy as a function of $b$ for $\frac{\pi}{4}<b<\frac{\pi}{2}$.}
\end{figure}

Notice that at the phase transition line $b=\frac{\pi}{4}$, $|E_b|=T_k$ and $E_b$, which is the boundary gap, sets the scale for the problem in this regime.

The density of continuous real root distribution in the ground state is given by
\begin{equation}
    2\rho(\lambda)=\frac{\text{sech}(\lambda )}{\pi } \left(2 \cosh ^2(\lambda ) \left(\frac{\cos (2 b)}{\cos (4 b)+\cosh (2 \lambda )}\right)+N+1\right)-\delta(\lambda).
\end{equation}

Hence, the the state described by this continuous distribution of all the real root is
\begin{equation}
    E_{ar}=\int\mathrm{d}\lambda\rho(\lambda)\frac{2J}{\cosh(\lambda)}=-\frac{2J(N+1)}{\pi}-\frac{2 J \tan ^{-1}(\tan (2 b)) \csc (2 b)}{\pi }+J.
\end{equation}
The ground state contains the discrete imaginary root $\lambda_b$ with energy $E_b$ since this solution has negative energy. Adding the energy of the boundary string solution, we obtain the ground state energy as 
\begin{equation}
   E_{gs}=E_b+E_{ar}= -\frac{2 J (N+1)}{\pi }-\frac{4 b J \csc (2 b)}{\pi }+J.
   \label{gsbmeng}
\end{equation}
Notice that the functional form of the equation is same as that of the ground state in the Kondo phase given by Eq.\eqref{gsK} which shows that the energy is continuous across the phase boundary at $b=\frac{\pi}{4}$.

Notice that at the phase boundary $b=\frac{\pi}{4}$, the last two terms in Eq.\eqref{gsbmeng} cancel and hence
\begin{equation}
    E_{\ket{gs}}\left(b=\frac{\pi}{4}\right)=-\frac{2J}{\pi}(N+1),
\end{equation}
which is equal to the energy of a periodic $XX$ chain with $N+1$ site. This is the only point in the entire phase space where the energy density is independent of the system size as mentioned earlier.

The bound mode is an exponentially localized mode with support around the impurity site. The wavefunction $F_b(j)$ for the bound mode can be obtained using $\lambda_b=\frac{i}{2}(4b-\pi)$ in Eq.\eqref{wfn}, and properly normalizing which gives 
\begin{equation}
    F_b(j)=\frac{2\gamma(b,j) \left(\tan^{j-N-1}(b)-\tan^{-j+N+1}(b)\right)}{\sqrt{\sec (2 b) \tan ^{-2 (N+1)}(b) \left(3-8 \sin ^4(b) \tan ^{4 N+2}(b)+\cos (4 b)\right)-4\cos (2 b)-8 (N+1)}},\nonumber
\end{equation}
where $\gamma(b,0)=\cos(b)$ and  $\gamma(b,j\neq 0)=1$. Upon taking the thermodynamic limit $N\to \infty$, the normalized wavefunction can be written as
\begin{equation}
    F_b(j)=\frac{2 \tan ^{-j}(b) \gamma (b,j)}{\sqrt{(\cos (4 b)-1) \sec (2 b)}}.
    \label{wnfterm}
\end{equation}

\begin{figure}[H]
    \centering
    \includegraphics[scale=0.5]{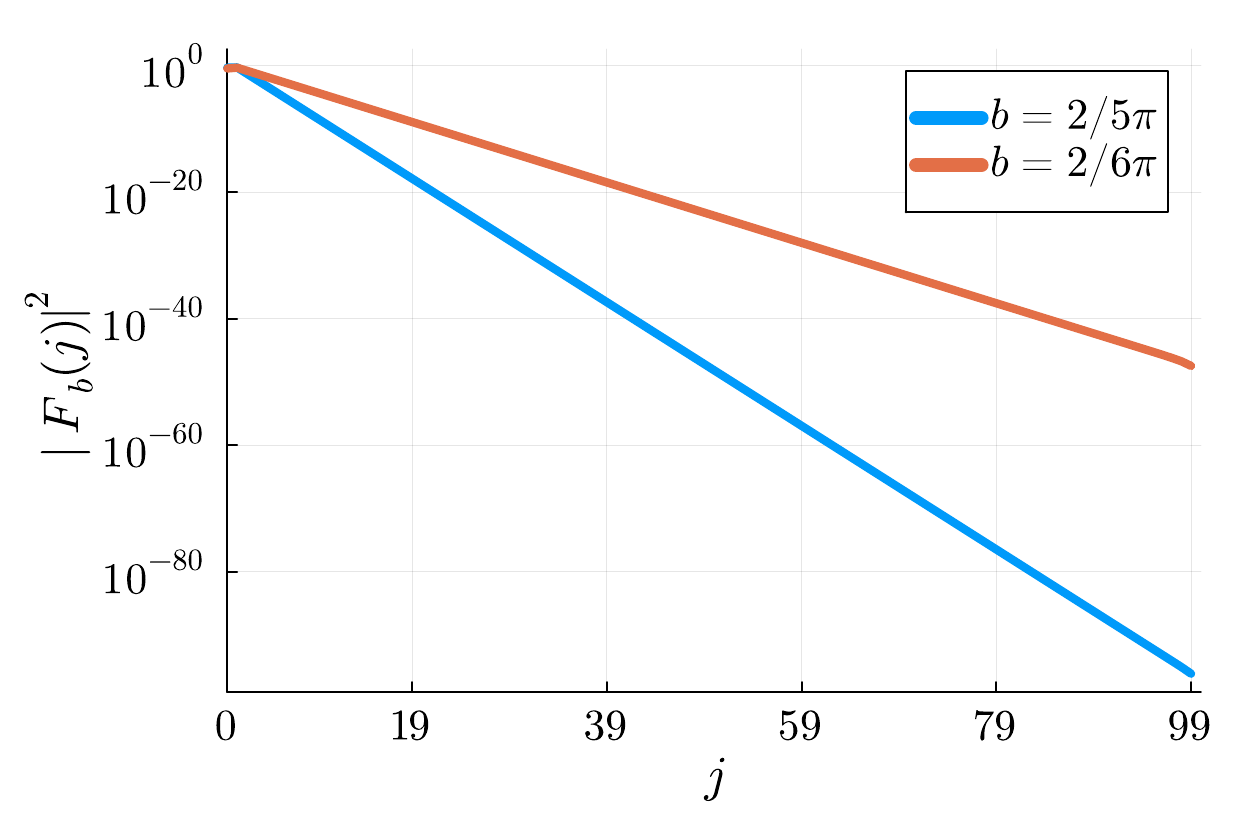}
    \caption{The modulus of the bound mode wavefunction localized at the left edge for different values of impurity coupling strength $b$ given by Eq.\eqref{wnfterm}. Here, $0$ is the impurity site and $j\in [1,N]$ are the bulk sites. In bulk, the impurity wavefunction falls off exponentially.}
    \label{fig:wfn}
\end{figure}
Using the relation \eqref{bstringeng}, we write the wavefunction in terms of the boundary string energy, the fundamental scale in the bound mode phase as
\begin{equation}
    F_b(j)=2^{j-\frac{1}{2}} \sqrt{E_b} J^{j-1} \sqrt[4]{E_b^2-4 J^2} \left(E_b+\sqrt{E_b^2-4 J^2}\right)^{-j}\upsilon(b,j),
\end{equation}
where $\upsilon(b,0)=\cos \left(\frac{1}{2} \sin ^{-1}\left(\frac{2 J}{E_b}\right)\right)$ and $\upsilon(b,j\neq 0)=1$.  This allows us to determine the localization length of the bound mode through writing $F_b(j)\sim e^{-j/\xi_b}$, which yields 
\begin{equation}
    \frac{1}{\xi_b}=\log\left((E_b/J)+\sqrt{(E_b/J)^2-4})\right).
\end{equation}

In this phase, apart from the bulk excitations that are constructed by adding spinons, boundary excitations are also possible. The boundary excitations are constructed by removing the boundary string solutions.  For example, we could remove the boundary string from the ground state and add a hole to construct a four-fold degenerate state with energy
\begin{equation}
    E_{\ket{ar,\theta}}=E_{ar}+E_\theta=-\frac{2 J (N+1)}{\pi }+J+ \frac{2J}{\cosh(\theta)}.
\end{equation}
The impurity is unscreened in this phase and hence, it can make singlet or triplet pairing with the spinon. In the thermodynamic limit, the singlet and triplet has the same energy. Thus, this is a four-fold degenerate state.

Starting from either the ground state $\ket{gs}$ or the state $\ket{ar,\theta}$, two distinct towers of the excited state can be built by adding an even number of spinons. The first tower built on top of the ground state $\ket{gs}$ contains all excited states in which the impurity is screened by the bound mode formed at the impurity site. However, the second tower built on top of $\ket{ar,\theta}$ contains all the states in which impurity is not screened. Notice that this is in sharp contrast with the Kondo phase where the impurity is screened in all eigenstates at zero field and zero temperature. 

Due to the presence of the isolated imaginary root, the ratio of the boundary and bulk contribution to the spinon density of states becomes
\begin{equation}
    R_{b}(E)=R(E)+\delta(E-E_b),
    \label{impdosbm}
\end{equation}
where $R(E)$ given by Eq.\eqref{redef} is negative in this phase and all the spectral weight comes from the bound mode thereby showing that impurity is screened by an exponentially localized bound mode formed at the impurity site. This qualitative difference in the observable like density of states in the Kondo and bound-mode phase shows that the boundary phase transition manifests itself in local physical quantities.\\

\subsubsection{\color{blue}{Effect of magnetic field}}\phantom{}\\

 In this subsection, we will compute the magnetization at the impurity site and show that the magnetization curve is qualitatively different compared to the Kondo phase. We  obtain the values of $k_j<k_F$ in the presence of magnetic field $H$ for a chain with $N=999$ and plot the magnetization in at the impurity site given by Eq.\eqref{magimpsite}.
 
\begin{figure}[H]
    \centering
    \includegraphics[scale=0.45]{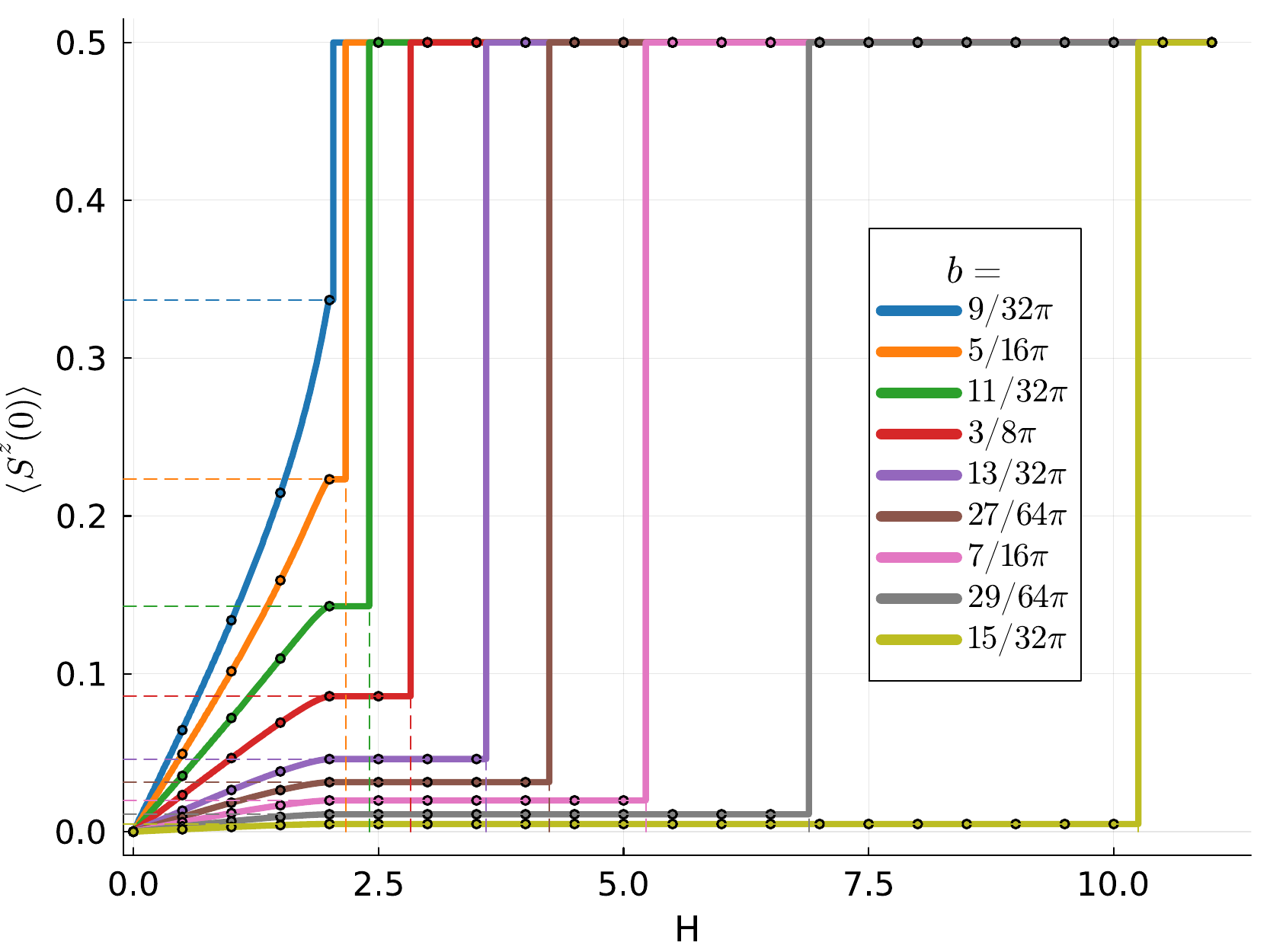}
    \caption{Impurity magnetization in the bound mode phase for $N=999$. The magnetization grows smoothly up to $H_c=2J$, then it saturates to a finite value $S^z_c(0) =\frac{\cot ^2(b)}{2}$ (shown in the dashed horizontal lines) before abruptly jumping to $\frac{1}{2}$  when the applied magnetic field $H$ is equal to the energy of the bound mode $|E_b|=2J\csc(2b)$ (shown in the dashed vertical lines).}
    \label{fig:pltbm}
\end{figure}

 Notice that only one $\lambda$ and hence one $k$ is complex. Thus, there is only one single particle mode that has an energy ($-2J\cos(k)$) greater than the maximum energy of the spinon $2J$ while the rest have an energy between $0<E<2J$. Thus, when a global magnetic field $H$ is applied, the magnetization at the impurity given by $\langle S^z(0)\rangle=\sum_{k<k_F(H)}S^z_k(0)$
 grows smoothly between $H=0$ and $H_c=2J$ where there are propagating spinons. However, when the magnetic field surpasses the critical field, there are no more spinons available to polarize. However, there is yet an unpolarized bound mode with energy $E_b$. Thus, the magnetization at the impurity site is constant between $H_c=2J$ and $H=E_b$ and exactly at $H=E_b$, the bound mode polarizes and hence the magnetization at the impurity site abruptly jumps to $\frac{1}{2}$ as shown in Fig.\ref{fig:pltbm}.
The value of $S^z(0)$ at the plateau between $H_c<H<E_b$, which is the contribution from all real $\lambda$ roots of Bethe equation, can be analytically computed for different values of $b$  as
\begin{equation}
  S^z_c(0)= F_{k_b}^*(0)F_{k_b}(0),
  \label{platval}
\end{equation}
where $k_b$ is the one imaginary $k$ solution which, in the thermodynamic limit, is given as
\begin{equation}
    k_b=i \log (\tan (b)).
\end{equation}

Thus, in the thermodynamic limit, the value at the plateau given by Eq.\eqref{platval} becomes
\begin{equation}
  S^z_c(0) =\frac{\cot ^2(b)}{2}
\end{equation}
and the contribution to the magnetization at the impurity site due to single imaginary $k$ solution which measures the height of the vertical jump $b_h$ in Fig.\ref{fig:pltbm} is
\begin{equation}
    b_h=\frac{1}{2}- S^z_c(0)=1-\frac{\csc ^2(b)}{2}.
    \label{verjump}
\end{equation}

\begin{figure}[H]
    \centering
    \includegraphics[scale=0.55]{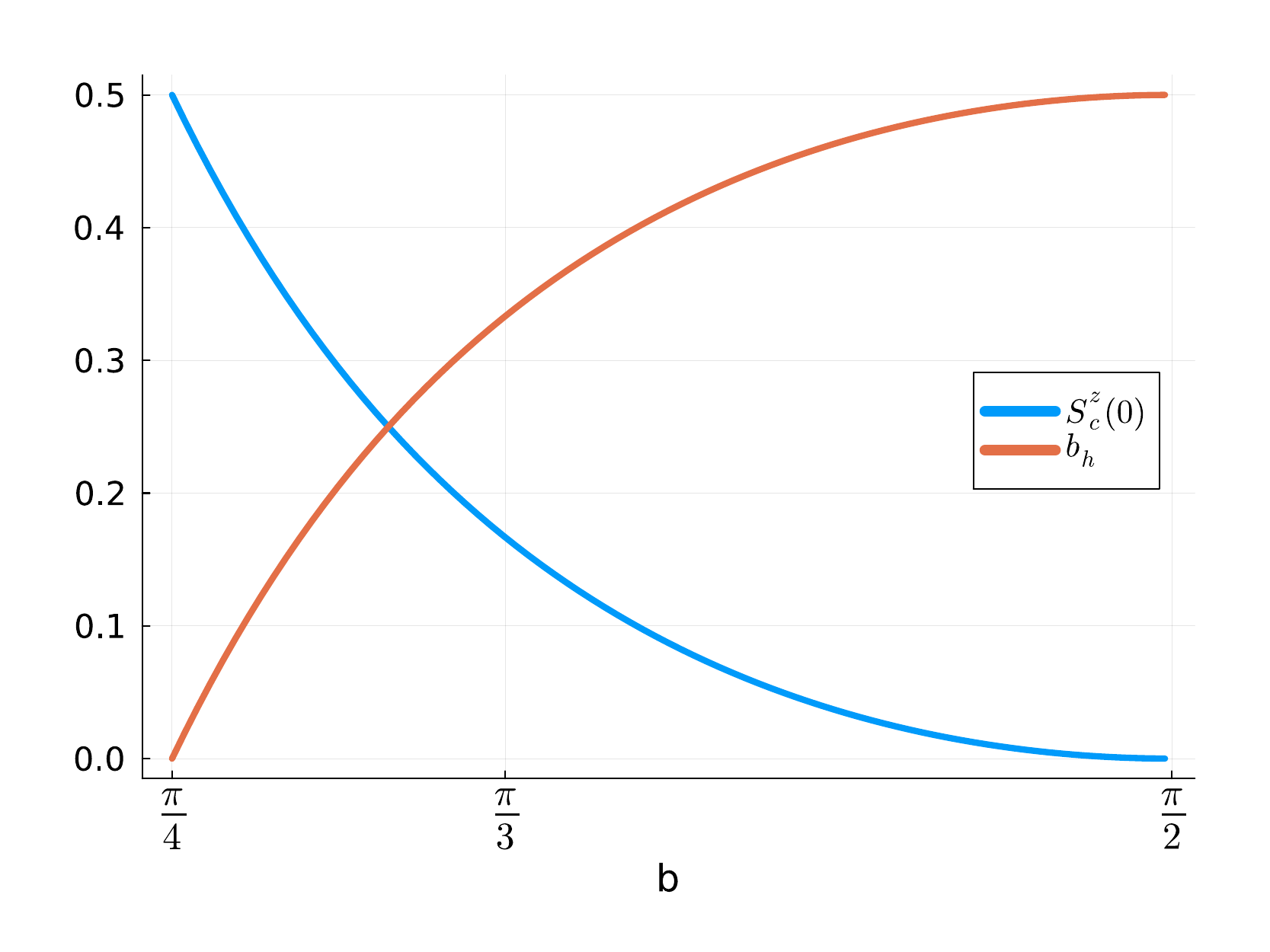}
    \caption{The plot of the magnetization at the impurity site contribution due to the all real roots of the Bethe equation ($  S^z_c(0)$) given by Eq.\eqref{platval}
and the contribution from the lone complex root ($b_h$) given by Eq.\eqref{verjump}.}
    \label{fig:jump}
\end{figure}

All the contribution to magnetization at the impurity site comes from the real roots of the Bethe equation at the phase boundary $\frac{\pi}{4}$ where $b_h$ is $0$ , at $b=\frac{\pi}{2}$, all the contribution comes from the complex root. At $b=\tan ^{-1}\left(\sqrt{2}\right)$, the contribution from $N-1$ real roots and $1$ complex root is equal as depicted in Fig. \ref{fig:jump}.

\begin{figure}
  \begin{minipage}{0.5\textwidth}
    \centering
    \includegraphics[width=\linewidth]{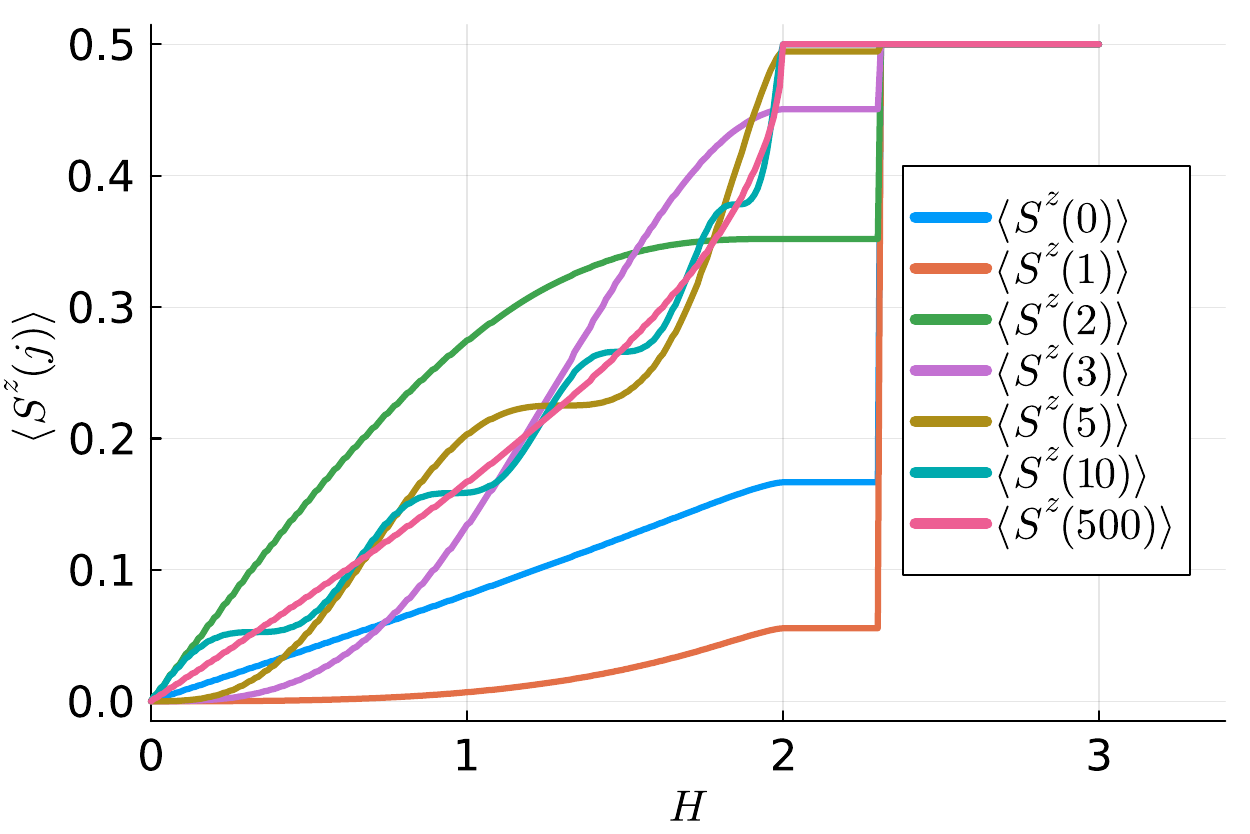}
    \label{fig:sub1}
  \end{minipage}%
  \begin{minipage}{0.5\textwidth}
    \centering
    \includegraphics[width=\linewidth]{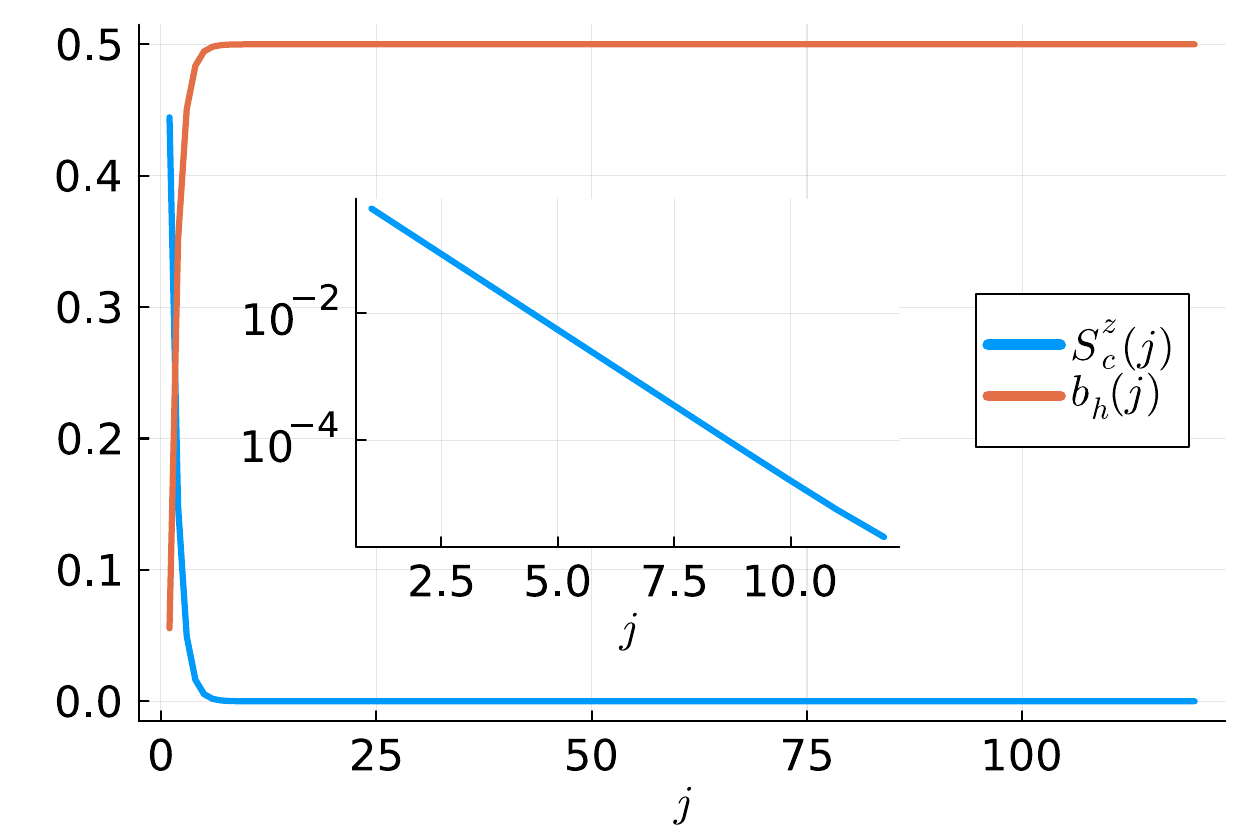}
    \label{fig:sub2}
  \end{minipage}
  \caption{The plot on the left shows the magnetization $S^z_c(j)$ at various sites for impurity parameter $b=\frac{2}{6}\pi$ given by Eq.\eqref{szcjeqn}. The magnetization jumps only for a few initial sites at the left edge of the chain. For all other sites, the magnetization grows continuously and reaches $\frac{1}{2}$ exactly at $H=2J$. The plot on the right shows the contribution to the magnetization from all real roots of the Bethe equation $S^z_c(j)$ and the single complex root $b_h(j)$ at various sites for $b=\frac{2}{6}\pi$. The inset shows the exponential falloff of the $S^z_c(j)$. The Bethe equation's real root solutions dominate magnetization contributions deep in the bulk. Although the bound mode affects initial few bulk sites due to exponential localization at the left edge, the magnetization in subsequent sites has a contribution solely from the real roots.}
  \label{fig:szpltbulk}
\end{figure}

The bound mode is localized at the left edge of the chain as shown in Fig.\eqref{fig:wfn}. Hence, the jump in magnetization also happens only at the sites close to the left edge. For a chain of $N=999$ bulk sites, we explicitly compute 1000 values of $k_j$ by solving the transcendental equation Eq.\eqref{kquant} for $b=\frac{2}{6}\pi$ and compute the magnetization at various sites in the presence of a global magnetic field $H$. As shown in Fig.\ref{fig:szpltbulk}, the magnetization jumps only at a few sites on the left and of the chain. For every site other than the first few sites, the magnetization reaches $\frac{1}{2}$ at $H=2J$ continuously. However, for the initial few sites, due to the presence of the localized bound mode, the magnetization reaches a finite value less than $\frac{1}{2}$ at $H=2J$ and plateaus until $H=E_b$ where it abruptly jumps to $\frac{1}{2}$.
We can compute the contribution to the magnetization due to all real root solutions of the Bethe equations for all the bulk sites in the thermodynamic limit
\begin{equation}
     S^z_c(j)= F_{k_b}^*(j)F_{k_b}(j)=\frac{\cos (4 b)-8 \cos (2 b) \cot ^{2 j}(b)-1}{2 (\cos (4 b)-1)}.
     \label{szcjeqn}
\end{equation}
Likewise, the value of the vertical jump at each site can be computed in the thermodynamic limit as
\begin{equation}
    b_h(j)=\frac{1}{2}- S^z_c(j)=\frac{4 \cos (2 b) \cot ^{2 j}(b)}{\cos (4 b)-1}.
\end{equation}
Clearly the function $S^z_c(j)$ exponentially increases to $\frac{1}{2}$ as $j$ increases as shown in Fig.\ref{fig:szpltbulk}. This shows that deep in the bulk all the contribution to the magnetization comes from the real roots of the Bethe equation. In the few initial sites, the contribution from the bound mode $ b_h(j)$ is significant but then it quickly approaches zero as $j$ increases. 

\subsubsection{\color{blue}{{Finite temperature effects}}}\phantom{}\\

The susceptibility of the impurity  in this phase takes the form
\begin{equation}
    \chi_{\mathrm{imp}}(T)=\int_0^\pi \frac{ \cos (2 b) \text{sech}^2\left(\frac{J\cos (k)}{T}\right)}{\pi  T \left(1-\sin ^2(2 b) \cos ^2(k)\right)}\mathrm{d}k+\frac{2\text{sech}^2\left(\frac{J \csc (2 b)}{T}\right)}{T},
    \label{susbmp}
\end{equation}
where the first term is the impurity contribution from the continuous real root distribution just like in the Kondo phase (ref Eq.\eqref{susint}) and the second term is the contribution from explicitly adding the isolated complex root of the Bethe Ansatz equations.
\begin{figure}[H]
    \centering
    \includegraphics[scale=0.65]{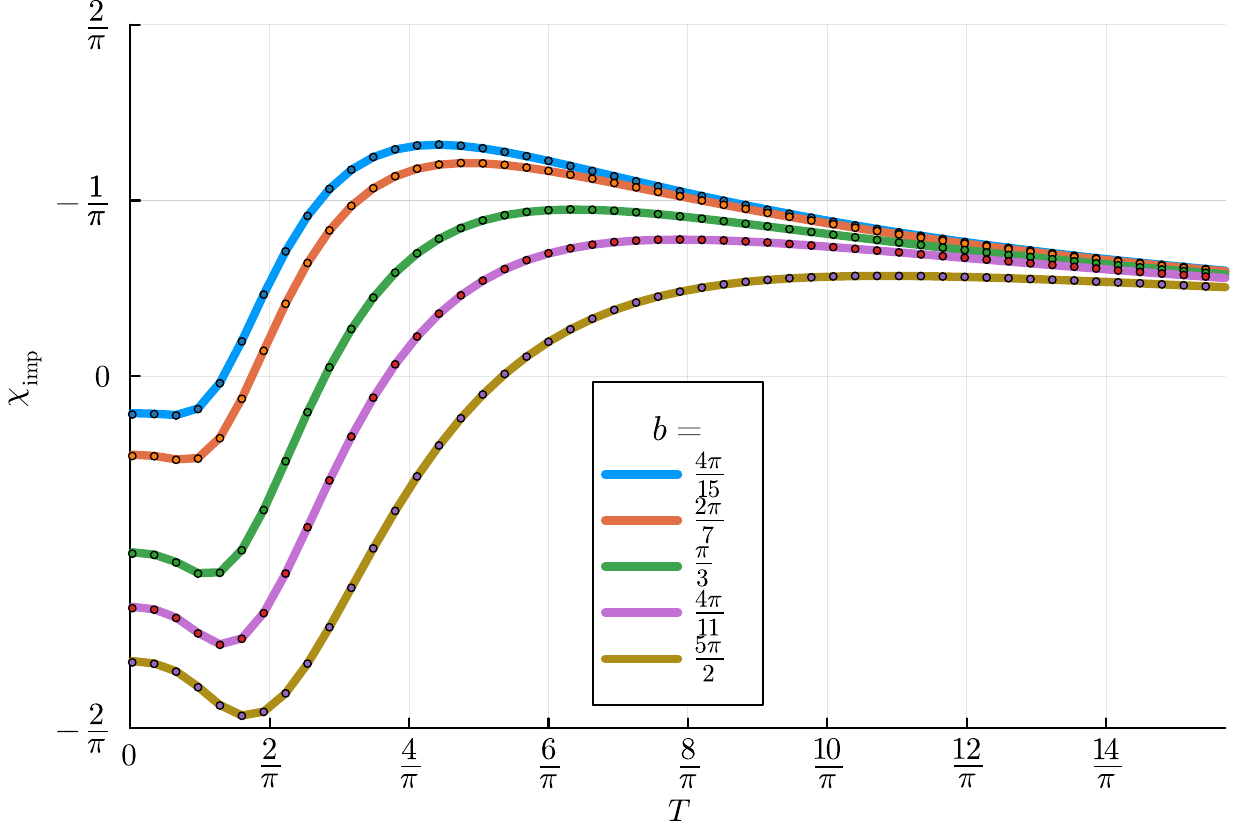}
    \caption{At zero temperature, the susceptibility in the bound mode phase is finite and negative:  -$\frac{2}{\pi}\cos(2b)$. As the temperature increases, the susceptibility first decreases and then increases to a positive value and attains a maximum. After that, the susceptibility falls off as $\frac{1}{T}$ at high energy showing that the impurity behaves as free spin at high temperature. The solid line is obtained by plotting Eq.\eqref{susbmp} whereas the discrete values shown in the plot are obtained from exact diagonalization by computing the impurity susceptibility  upon subtracting the susceptibility of the free chain with 499 site containing no impurity from the susceptibility of a chain with 500 sites with the impurity at its edge.}
    \label{fig:susbm}
\end{figure}
At low temperature, the asymptotic value of susceptibility is 
\begin{align}
    {\chi_{\mathrm{imp}}}_{(T\to 0)}&=\frac{2 \cos (2 b) \tanh \left(\frac{J}{T}\right)}{\pi  J}+\frac{2\text{sech}^2\left(\frac{J \csc (2 b)}{T}\right)}{T}+\cdots\nonumber\\
    &=\frac{2 \text{sech}\left(\frac{E_{\text{b}}}{2 T}\right)}{T}-\frac{2 \sqrt{1-\frac{4 J^2}{E_{\text{b}}^2}} \tanh \left(\frac{J}{T}\right)}{\pi  J}+\cdots
\end{align}
where $\cdots$ represent the higher order correction terms that vanish when $T\to 0$. Note that at $T=0$, the susceptibility of impurities becomes $\frac{2}{\pi}\cos(2b)$ which is negative in the bound mode phase \textit{i.e.} when $\frac{\pi}{4}<b<\frac{\pi}{2}$. We interpret the negative susceptibility as a result of the bound mode acting as a local magnetic field in the opposite direction of the external magnetic field due to the nature of the impurity tending to form a singlet with the bound state. Similar physics was also observed in the Kondo impurity in a superconducting wire\cite{pasnoori2020kondo}.

As the temperature increases, the susceptibility first decreases, attains a minimum value, and then starts to increase to become positive. After it attains a maximum positive value at some finite temperature, it starts to fall off as $\frac{1}{T}$ demonstrating that the impurity behaves as free spin at high temperature as shown in Fig.\ref{fig:susbm}. The high-temperature asymptotic expansion of susceptibility can be written as
\begin{equation}
   {\chi_{\mathrm{imp}}}_{(T\to \infty)}=  \frac{1}{T}\left(1-\frac{J^2 \sec ^2(b)}{2 T^2}\right)+\mathcal{O}\left(\frac{1}{T^4}\right)=\frac{1}{T}-\frac{E_{\text{b}}^2 \left(\sqrt{1-\frac{4 J^2}{E_{\text{b}}^2}}+1\right)}{4 T^3}+\mathcal{O}\left(\frac{1}{T^4}\right)
\end{equation}
where we used Eq.\eqref{bstringeng} at the last step.

\section{Conclusion}
We summarize the key aspect of our work. Considering the spin-$\frac{1}{2}$ $XX$ chain with boundary impurity which is equivalent to the lattice version of the spin sector of conventional Kondo problem in the low energy regime, we analyze it analytically using \textit{Bethe Ansatz} as well as numerically using \textit{exact diagonalization}.  We found that the boundary phenomena depend on the ratio of the boundary couplings to the bulk coupling. We showed that the model exhibits two distinct kinds of phase: the Kondo phase, which is characterized by screening of the impurity by the multi-particle Kondo cloud, and the bound mode phase, which is characterized by screening of the impurity spin by a single-particle bound mode formed at the impurity site. The signature of the phase transition is seen in several physical quantities such as the spinon density of states. impurity magnetization and susceptibility. The impurity density of states changes from the characteristic Kondo peak at $E=0$ to a peak at $E=2 J $ and eventually becomes a delta function peak at $E_b$ in the bound mode regime.  Using both Bethe Ansatz and exact diagonalization we showed that the magnetization at the impurity site in the Kondo phase shows a smooth crossover from 0 to $\frac{1}{2}$ just as in the Fermi liquid Kondo. However, in the bound mode phase, the magnetization undergoes a sudden jump from some finite value to $\frac{1}{2}$ as the magnetic field increases. The sudden jump of impurity magnetization occurs when the external magnetic field is equal to the energy of the bound mode. Likewise, the finite temperature susceptibility behaves differently in these two phases. In the Kondo phase, the susceptibility is finite and positive at zero temperature and asymptotically free exhibiting Curie law behavior in the high temperature regime just like in the conventional Kondo problem. However, due to the competition between the local bound mode formed at the impurity site and the applied global magnetic field, the impurity susceptibility is finite but negative at zero temperature. Upon increasing the temperature, the susceptibility first decreases and then starts to increase and attains some finite maximum value. Eventually, the susceptibility falls off as $\frac{1}{T}$ demonstrating the Curie law which shows that at high temperature, the impurity is essentially free. 

Across the Kondo-bound mode phase boundary, there is a distinctive change in the nature of the ground state. In the Kondo phase, the ground state (and all other states built on it) hosts an impurity that is screened by a many body Kondo cloud whereas in the bound mode phase the screening shifts to a single-particle effect. Moreover, the entire structure of the Hilbert space reorganizes into two distinct towers of excited state in the bound mode phase: one containing all the states where impurity is screened by bound mode, and the other one containing all the states where impurity is unscreened. This phenomenon of the change in the number of towers of  called `boundary eigenstate phase transition' is observed in other one-dimensional models \cite{pasnoorithesis,pasnoori2023boundary, kattel2023kondo}. Our study demonstrates that these boundary phase transitions reflect in local observables, such as the impurity density of states and local magnetization at the impurity site, exhibiting distinct behaviors in the Kondo and bound mode phases.

 The interacting case of $XXX$ model with $XXX$ impurity was studied in \cite{kattel2023kondo} and it was shown that the boundary phase transition occurs between the Kondo phase and the bound mode phase when the boundary and bulk coupling ratio is $J_{\mathrm{imp}}/J=\frac{4}{3}$. Here, we showed that all of the essential boundary features reported there exist in the this case also. Thus, the $\sigma^z_j\sigma^z_{j+1}$ interaction seems to only play a role in dressing up the bare parameter like in Fermi liquid theory. 

The implication of the existence of boundary bound mode in the dynamics of the model is an important question. In an upcoming work, we study the non-equilibrium aspect of the model focusing on quench dynamics and also the effect of the boundary eigenstate phase transition in the dynamics.

\section{Acknowledgement}
We thank Parameshwar R.~Pasnoori whose questions served as inspiration for this project. The helpful discussions with Colin Rylands are gratefully acknowledged. J.H.P. is partially supported by NSF Career Grant No.~DMR- 1941569 and the Alfred P.~Sloan Foundation through a Sloan Research Fellowship.

\section*{References}
\bibliographystyle{unsrt}
\bibliography{ref}

\begin{appendix}
\section{One particle wavefunction}\label{wfnsec}
Rewrite the Hamiltonian Eq.\eqref{hammodel} as
\begin{equation}
    H= \sum_{j=1}^{\mathcal{N}-1} J(\sigma^+_j\sigma^-_{j+1}+\sigma^+_{j+1}\sigma^-_j)+ J_\mathrm{imp} (\sigma_1^+\sigma_0^-+\sigma_0^+\sigma_1^-)
\end{equation}
using the relation $\sigma_j^{ \pm}=\frac{1}{2}\left(\sigma_j^x \pm i \sigma_j^y\right)$.

    The wavefunction in the one particle sector can be written as
    \begin{equation}
        \sum_{j=0}^N F(j)\ket{\uparrow_0,\uparrow_1, \cdots, \uparrow_{(j-1)}, \downarrow_j, \uparrow_{j+1}, \cdots, \uparrow_N}
    \end{equation}

    For $j\neq 0,1$ and ${N}$, the schrodinger's equation becomes
    \begin{equation}
        J F(j-1)+J F(j+1)=E F(j)
    \end{equation}
    Proposing the wavefunction of the form
    \begin{equation}
        F(j)=A(k)e^{-ikj}+B(k)e^{ikj}+C(k)\delta_{j,0}
    \end{equation}
    we obtain the energy to be
    \begin{equation}
        E=2 J\cos k
    \end{equation}
    and the wavefunction at $j=0,1$ and $N$ can be written as \begin{align}
       J_\mathrm{imp} F(1)&= 2J\cos k F(0)\\
       J F({N}-1) &= 2J\cos k F({N})\\
      J_\mathrm{imp} F(0)+ J F(2) &=2J\cos k F(1)
    \end{align}

    Solving these equations, we obtain the quantization condition
    \begin{equation}
        2 J^2 \cos (k) \sin (k {N})-{J_\mathrm{imp}}^2 \sin (k ({N}-1))=0
        \label{quantcond}
    \end{equation}

    Such that the non-normalized wavefunction takes the form
    \begin{equation}
        F(j)=\begin{cases}
            J \sin (k (N+1)) \quad \quad \quad \quad \quad ~ \text{when}\quad j=0\\ 
            J_\mathrm{imp} \sin (k ({N}+1-j)) \quad \text{when}\quad j\neq 0
        \end{cases}
        \label{wfn}
        \end{equation}

 Notice that the quantization condition Eq.\eqref{quantcond} is just the Bethe equation Eq.\eqref{baebae}. To prove this, consider the Bethe equation

        \begin{equation}
            \left(\frac{\sinh \left(\frac{\lambda_j}{2}+\frac{i \pi}{4} \right)}{\sinh \left(\frac{\lambda_j}{2}-\frac{i \pi}{4} \right)}\right)^{2 N+2}\frac{\sinh \left(\frac{\lambda_j}{2}+i  b+\frac{i \pi}{4} \right)}{\sinh \left(\frac{\lambda_j}{2}-i b-\frac{i \pi}{4} \right)}  \frac{\sinh \left(\frac{\lambda_j}{2}-i  b+\frac{i \pi}{4} \right)}{\sinh \left(\frac{\lambda_j}{2}+i b-\frac{i \pi}{4} \right)}   =1
        \end{equation}

        and perform a change of the variable
        \begin{equation}
            \frac{\sinh \left(\frac{\lambda_j}{2}+\frac{i \pi}{4} \right)}{\sinh \left(\frac{\lambda_j}{2}-\frac{i \pi}{4} \right)}=e^{-ik_j}
        \end{equation}
        such that upon using $J_{\mathrm{imp}}=J \sec(b)$, one obtains
        \begin{equation}
            e^{-2ik_j(N+1)}\frac{-e^{2 i k_j} \left(J_\mathrm{imp}\right)^2+J^2 e^{2 i k_j}+J^2}{-\left(J_\mathrm{imp}\right)^2+J^2 \left(1+e^{2 i k_j}\right)}=1
        \end{equation}
        which can be written as
        \begin{equation}
            \left(J_\mathrm{imp}\right)^2 \sin (k N)-2 J^2 \cos (k) \sin (k (N+1))=0
        \end{equation}
        which is exactly the same as the quantization condition obtained in \ref{quantcond}.

        Moreover, the energy relation Eq.\eqref{engrel} becomes \begin{equation}
            E=2J\sum_{k_j}\cos(k_j).
        \end{equation}
\end{appendix}

\end{document}